\documentclass[12pt,manuscript]{aastex}

\newcommand{\simgt} {\,\hbox{\lower0.6ex\hbox{$\sim$}\llap{\raise0.6ex\hbox{$>$}}}\,}
\newcommand{\simlt} {\,\hbox{\lower0.6ex\hbox{$\sim$}\llap{\raise0.6ex\hbox{$<$}}}\,}
\newcommand{\beq}{\begin{equation}}
\newcommand{\eeq}{\end{equation}}
\def\epsw{\epsilon_{\rm w}}
\def\epswM{\epsw^{\rm M}}
\def\etaw{\eta_{\rm w}}
\def\etawM{\etaw^{\rm M}}
\def\stacksymbols #1#2#3#4{\def\theguybelow{#2}
        \def\verticalposition{\lower#3pt}
        \def\spacingwithinsymbol{\baselineskip0pt\lineskip#4pt}
        \mathrel{\mathpalette\intermediary#1}}
\def\intermediary #1#2{\verticalposition\vbox{\spacingwithinsymbol
        \everycr={}\tabskip0pt
        \halign{$\mathsurround0pt#1\hfil##\hfil$\crcr#2\crcr
                \theguybelow\crcr}}}

\catcode`\@=11
\def\gsim{\ifmmode{\mathrel{\mathpalette\@versim>}}
    \else{$\mathrel{\mathpalette\@versim>}$}\fi}
\def\lsim{\ifmmode{\mathrel{\mathpalette\@versim<}}
    \else{$\mathrel{\mathpalette\@versim<}$}\fi}
\def\@versim#1#2{\lower 2.9truept \vbox{\baselineskip 0pt \lineskip 
    0.5truept \ialign{$\m@th#1\hfil##\hfil$\crcr#2\crcr\sim\crcr}}}
\catcode`\@=12

\def\pd#1#2{\partial #1\over {\partial #2}}
\def\brem{bremsstrahlung$\;\,$}

\def\Msun{M_{\odot}}
\def\kb{k_{\rm B}}
\def\mpr{m_{\rm p}}
\def\eps{\epsilon}
\def\epsz{\epsilon_0}
\def\epsII{\eps_{\rm II}}
\def\epsw{\eps_{\rm w}}
\def\epswM{\epsw^{\rm M}}
\def\epsj{\eps_{\rm j}}
\def\epsopt{\eps_{\rm opt}}
\def\epsUV{\eps_{\rm UV}}
\def\epsA{\eps_{\rm EM}}
\def\tauII{\tau_{\rm II}}
\def\tauopt{\tau_{\rm opt}}
\def\tauUV{\tau_{\rm UV}}
\def\tform{\tau_{\rm form}}
\def\tdyn{\tau_{\rm dyn}}
\def\tcool{\tau_{\rm cool}}
\def\tjeans{\tau_{\rm Jeans}}
\def\trot{\tau_{\rm rot}}

\def\taul{\tau_{\rm *l}}
\def\tauh{\tau_{\rm *h}}
\def\tlagd{\tau_{\rm d}}
\def\tlagi{\tau_{\rm i}}

\def\lb{L_{\rm B}}
\def\lbh{L_{\rm BH}}

\def\lsn{L_{\rm SN}}

\def\ledd{L_{\rm Edd}}
\def\ldwin{L_{\rm dw}}
\def\lj{L_{\rm j}}
\def\luv{L_{\rm UV}}
\def\lopt{L_{\rm opt}}
\def\lbhefUV{L_{\rm BH,UV}^{\rm eff}}
\def\lbhefopt{L_{\rm BH,opt}^{\rm eff}}
\def\lbhefphot{L_{\rm BH,photo}^{\rm eff}}
\def\lbhef{\lbh^{\rm eff}}
\def\luveff{\luv^{\rm eff}}
\def\lopteff{\lopt^{\rm eff}}
\def\lir{L_{\rm IR}}
\def\lduv{L_{\rm d,UV}}
\def\ldopt{L_{\rm d,opt}}
\def\lrad{L_{\rm r}}
\def\Pwj{P_{\rm wj}}
\def\Ywj{Y_{\rm wj}}
\def\mast{M_*}

\def\Min{M_{\rm inf}}
\def\MII{M_{\rm II}}
\def\MTO{M_{\rm TO}}
\def\mbh{M_{\rm BH}}

\def\Mdg{M_{\rm dg}}
\def\Mds{M_{\rm d*}}
\def\Mdsl{M_{\rm dl*}}
\def\Mdsh{M_{\rm dh*}}
\def\Mdw{M_{\rm dw}}
\def\Mrem{M_{\rm rem}}
\def\Mfid{M_{\rm fid}}
\def\Medd{M_{\rm Edd}}
\def\Mj{M_{\rm j}}

\def\mdot{\dot\mbh}
\def\dmin{\dot M_{1}}
\def\dmineff{\dot M_1^{\rm eff}}
\def\drhoII{\dot\rho_{\rm II}}
\def\als{\alpha_*}
\def\asn{\alpha_{\rm SN}}
\def\Rsn{R_{\rm SN}}
\def\dMfid{\dot\Mfid}
\def\rhos{\rho_*}

\def\rx{R_{\rm X}}

\def\Rd{R_{\rm d}}

\def\kes{\kappa_{\rm es}}
\def\kUV{\kappa_{\rm UV}}
\def\kopt{\kappa_{\rm opt}}
\def\kpho{\kappa_{\rm photo}}
\def\kIR{\kappa_{\rm IR}}
\def\tx{T_{\rm X}}

\def\fh{f_{\rm h}}

\def\freml{f_{\rm rem,l}}
\def\fremh{f_{\rm rem,h}}
\def\etaII{\eta_{\rm SN}}
\def\etaD{\eta_{\rm d}}
\def\etaw{\eta_{\rm w}}
\def\etawM{\etaw^{\rm M}}
\def\etaj{\eta_{\rm j}}
\def\etas{\eta_*}
\def\etaform{\eta_{\rm form}}

\def\Esn{E_{\rm SN}}
\def\NII{N_{\rm II}}
\def\dEII{\dot E_{\rm II}}
\def\dEI{\dot E_{\rm I}}

\def\dEopt{\dot E_{\rm opt}}
\def\dEUV{\dot E_{\rm UV}}
\def\Juveff{J_{\rm UV}^{\rm eff}}
\def\Jopteff{J_{\rm opt}^{\rm eff}}

\def\vw{v_{\rm w}}
\def\vj{v_{\rm j}}
\def\vwj{v_{\rm wj}}
\def\vff{v_{\rm ff}}
\def\DOmew{\Delta\Omega_{\rm w}}
\def\DOmej{\Delta\Omega_{\rm j}}

\def\mw{m_{\rm w}}
\def\mj{m_{\rm j}}

\def\Pism{P_{\rm ISM}}

\def\sigast{\sigma_*}

\def\drhosp{\dot\rhos^+}
\def\prad{p_{\rm rad}}

\def\t15{t_{15}}
\def\vtsn{\vartheta_{\rm SN}}

\def\pd#1#2{\partial #1\over {\partial #2}}
\def\brem{bremsstrahlung$\;\,$}

\def\Msun{M_{\odot}}
\def\kb{k_{\rm B}}
\def\mpr{m_{\rm p}}
\def\eps{\epsilon}
\def\epsII{\eps_{\rm II}}

\def\epsopt{\eps_{\rm opt}}
\def\epsUV{\eps_{\rm UV}}
\def\tauII{\tau_{\rm II}}
\def\tauopt{\tau_{\rm opt}}
\def\tauUV{\tau_{\rm UV}}
\def\tform{\tau_{\rm form}}
\def\tdyn{\tau_{\rm dyn}}
\def\tcool{\tau_{\rm cool}}
\def\tjeans{\tau_{\rm Jeans}}
\def\trot{\tau_{\rm rot}}

\def\taul{\tau_{\rm *l}}
\def\tauh{\tau_{\rm *h}}
\def\lb{L_{\rm B}}
\def\lbh{L_{\rm BH}}

\def\lsn{L_{\rm SN}}

\def\ledd{L_{\rm Edd}}
\def\ldwin{L_{\rm dw}}
\def\luv{L_{\rm UV}}
\def\lopt{L_{\rm opt}}
\def\lbhefUV{L_{\rm BH,UV}^{\rm eff}}
\def\lbhefopt{L_{\rm BH,opt}^{\rm eff}}
\def\lbhefphot{L_{\rm BH,photo}^{\rm eff}}
\def\lbhef{\lbh^{\rm eff}}
\def\luveff{\luv^{\rm eff}}
\def\lopteff{\lopt^{\rm eff}}
\def\lir{L_{\rm IR}}
\def\lduv{L_{\rm d,UV}}
\def\ldopt{L_{\rm d,opt}}
\def\lrad{L_{\rm r}}
\def\mast{M_*}

\def\Min{M_{\rm inf}}
\def\MII{M_{\rm II}}
\def\MTO{M_{\rm TO}}
\def\mbh{M_{\rm BH}}

\def\mw{m_{\rm w}}

\def\mdot{\dot\mbh}
\def\dmin{\dot M_{1}}
\def\drhoII{\dot\rho_{\rm II}}
\def\als{\alpha_*}
\def\asn{\alpha_{\rm SN}}
\def\Rsn{R_{\rm SN}}
\def\rhos{\rho_*}

\def\rx{R_{\rm X}}
\def\kes{\kappa_{\rm es}}
\def\kUV{\kappa_{\rm UV}}
\def\kopt{\kappa_{\rm opt}}
\def\kpho{\kappa_{\rm photo}}
\def\kIR{\kappa_{\rm IR}}
\def\tx{T_{\rm X}}

\def\fh{f_{\rm h}}

\def\freml{f_{\rm rem,l}}
\def\fremh{f_{\rm rem,h}}
\def\etaII{\eta_{\rm SN}}
\def\etaD{\eta_{\rm d}}
\def\etaw{\eta_{\rm w}}
\def\etas{\eta_*}
\def\etaform{\eta_{\rm form}}

\def\Esn{E_{\rm SN}}
\def\NII{N_{\rm II}}
\def\dEII{\dot E_{\rm II}}
\def\dEI{\dot E_{\rm I}}

\def\dEopt{\dot E_{\rm opt}}
\def\dEUV{\dot E_{\rm UV}}
\def\Juveff{J_{\rm UV}^{\rm eff}}
\def\Jopteff{J_{\rm opt}^{\rm eff}}

\def\vw{v_{\rm w}}

\def\sigast{\sigma_*}

\def\drhosp{\dot\rhos^+}
\def\prad{p_{\rm rad}}

\shorttitle{Purely mechanical AGN feedback models}
\shortauthors{Shin et al.}

\begin{document}

\slugcomment{Submitted to ApJ; resubmitted on Dec. 27th, 2009}

\title{Feedback from central black holes in elliptical galaxies. II: 
Can purely mechanical energy feedback models work?}

\author{Min-Su Shin}
\affil{Princeton University Observatory, Peyton Hall, Princeton, NJ 08544-1001}

\author{Jeremiah P. Ostriker}
\affil{Princeton University Observatory, Peyton Hall, Princeton, NJ 08544-1001}
\affil{Institute of Astronomy, University of Cambridge, 
Madingley Road, Cambridge CB3 0HA, UK}

\author{Luca Ciotti}
\affil{Department of Astronomy, University of Bologna, via Ranzani 1, 
I-40127, Bologna, Italy}

\begin{abstract}
By using high-resolution 1D hydrodynamical simulations, 
we investigate the effects of purely mechanical feedback from super massive black holes 
(SMBHs) in the evolution of elliptical galaxies 
for a broad range of feedback efficiencies and compare
the results to four major observational constraints. 
In particular, 
we focus on 1) the central black hole to stellar mass ratio of the host galaxy, 
2) the lifetime of the luminous quasar phase, 
3) the mass of stars formed in the host galaxy within the last Gyr, and 
4) the X-ray luminosity of the hot diffuse gas. As a result, 
we try to pin down the most successful range of mechanical feedback efficiencies. 
We find that while low feedback 
efficiencies result in too much growth of the SMBH, high efficiencies totally 
blow out the hot interstellar gas, and 
the models are characterized by very low thermal X-ray luminosity well 
below the observed range. 
The net lifetime of the quasar phase is strongly coupled to the 
mass ratio between SMBH and its host galaxy, 
while the X-ray luminosity is generally correlated to the recent star 
formation within the last Gyr. 
When considering the popularly adopted model of the constant feedback efficiency, the feedback 
energy deposited into the ambient medium should be more than 0.01\% of the SMBH accretion 
energy to be consistent with the SMBH mass to stellar mass ratio in the local universe. Yet, 
the X-ray luminosity of the hot gas favors about 0.005\% of the accretion energy as the 
mechanical AGN feedback energy. 
We conclude that the purely mechanical feedback mode is unlikely to be simultaneously 
compatible with all four observable tests, even allowing a broad range of feedback efficiencies, and 
that including both radiative and mechanical feedback together may be a solution to comply 
the observational constraints. In addition to the adopted observational constraints, 
our simulations also show that 
the ratio of SMBH growth rate over its current mass and the density and temperature 
distribution of hot gas can be useful 
observable diagnostics for AGN feedback efficiencies.
\end{abstract}


\section{Introduction}

The well established empirical correlations between the mass of supermassive black holes 
(SMBHs) and several properties of their host galaxies 
are providing new insights and also new problems for our 
understanding of massive galaxy evolution. 
In particular, the more massive SMBHs are hosted in 
the more massive bulges, galaxies, and dark matter halos \citep{kormendy95,magorrian98,
ferrarese00,gebhardt00,ferrarese02,yu02,marconi03,haring04}, and the shape of the light profile is related 
to the mass of the central SMBH \citep[e.g.][]{graham01}. 
Moreover, the co-evolution of 
SMBHs and host galaxies is also supported by the 
observational evidence of the proportionality 
between the galaxy star formation rate (SFR) 
and the mass accretion rate of the SMBHs over a broad range 
of redshifts \citep[e.g.][]{boyle98,haiman04,miller07,shankar09}.

This tight correlation between the stellar mass of the host spheroid and 
the SMBH mass has been tackled by various theoretical 
explanations that are based on self-regulation processes of 
feedback effects from the accreting SMBHs. When SMBHs are in an active phase (i.e. AGN), 
feedback via 
either the mass ejection by winds or jets, or alternatively the emitted radiation, 
regulates the mass accretion rate and the final SMBH mass 
\citep{rees84,krolik99,heckman08}. 
The feedback can be in the forms of radiatively or mechanically driven winds 
\citep[e.g.][]{silk98,fabian99,king03a,granato04,king05,murray05,fabian06}, 
of a turbulent energy transportation near the SMBHs \citep[e.g.][]{begelman05}, 
of radiative effects such as 
photoionization and Compton heating \citep[e.g.][]{sazonov05,ciotti07}, 
or of a blast wave \citep[e.g.][]{menci08}.

The idea of feedback from SMBHs also has given rise to various hypotheses 
that could answer important open questions in galaxy and galaxy cluster evolution. 
For example, significant star formation can be prevented by 
energy deposited from AGNs, which suppresses the 
supply of cold gas, and the further late star formation after the bulk of stars form early 
\citep[e.g.][]{croton06,cattaneo07,
khalatyan08,cattaneo09,kormendy09} and transforms blue star-forming galaxies to red galaxies 
\citep[e.g.][]{lagos08}. The AGN feedback may help us 
explain galaxy downsizing by making the AGN feedback processes depend on 
the mass of dark matter halo \citep[e.g.][]{scannapieco05}. Moreover, the feedback may 
alter energy and mass distribution of intergalactic medium in galaxy clusters \citep{
cavaliere02,scannapieco04,scannapieco05,thacker06,chandran07,chatterjee08}.

Independently of the specific model considered, 
the problem of the self-regulated growth of SMBHs and spheroids can be summarized as two 
questions about effects from AGN feedback. First, self-regulation requires that 
feedback effects have to be well timed responding to the growth of stellar mass. Suppose 
that there is a long time-lag either between the start of star formation and the onset of 
AGN feedback, or between the beginning of the feedback effects and heating surrounding gas, then  
the pace of star formation can be faster than the growth of SMBHs. This case would weaken 
the tight correlation between SMBH mass and bulge mass. 
Therefore, this {\it timing problem} is closely relevant 
to how SMBHs are fueled and when they ignite AGN feedback \citep{cavaliere83,shull83,shlosman90,quataert00,
kawaguchi03,narayan05,tan05,jogee06,konigl06,davies07,muller09}. Second, the impact of AGN feedback 
on host galaxies of SMBHs depends on how the gravitational accretion energy is converted to 
out-flowing mass and radiation from SMBHs, and is transported to the 
surrounding medium \citep{king03b,proga05,murray05,proga07,proga08,hardee08}. 
In some circumstances, the energy conversion 
might even cause the enhancement of star formation instead of suppressing star formation 
by allowing shock-induced star formation 
\citep[e.g.][]{silk05,ciotti07,pipino09}. 
Very efficient energy output and conversion will tend to inhibit the SMBH growth overly, and 
very inefficient coupling of the outflowing energy to the ambient gas would lead to SMBHs more 
massive even than those observed. This {\it energy conversion problem} 
coupled to the {\it timing problem} determines the efficiency of AGN feedback and the final 
properties of SMBHs and their host galaxies.

Therefore, the proper understanding of 
the {\it timing problem} and the {\it energy conversion problem} 
is critical to understand physics of AGN feedback in detail. 
As \citet{elvis06} emphasizes, the structure and 
physics of AGNs are entangled together, and it is difficult to find a direct connection 
between feedback from AGNs and other properties of galaxies without elucidating 
the complicated connection of feedback effects onto the host galaxies. 
Even though we can begin to ask questions related to AGN feedback on a large scale such 
as galaxy mergers, galaxy clusters, and cosmological evolution by using 
simple descriptions of feedback 
\citep{cattaneo05,hopkins05,kawata05,thacker06,monaco07,menci08,khalatyan08}, we 
cannot simplify the complex physics entangling the SMBHs and their host galaxies without 
the better understanding of involved physics, ranging from the small-scale physics of AGNs 
to the larger-scale physics of galaxies \citep{blandford01}. 

In the literature, feedback has been mainly considered in two aspects: 
radiative feedback and mechanical feedback. 
Radiative feedback is a consequence of radiative 
heating and radiation pressure by a strong radiation from AGNs, 
including photoionization and Compton heating 
\citep{ciotti01,sazonov04,sazonov05,fabian06,ciotti07}. And 
computing the effects of radiative feedback is straightforward since we directly 
observe the hard radiation from AGNs that causes the ambient heating: there are 
few uncertainties in computing this type of feedback. Yet, mechanical feedback is 
associated to winds and jets from accreting SMBHs \citep{tabor93,binney95,friaca98,veilleux05,
springel05a,konigl06}, and it can be very important although the 
strength of the effects is highly uncertain. 
Both feedback modes increase the internal energy of the interstellar medium (ISM) and may 
cause local and/or global outflows. But their 
properties and efficiencies are different because radiative feedback 
(in an optically thin medium) acts almost instantaneously, and is transported 
without mass loading over the entire galaxy. By any accounting the radiative energy 
emitted by the accreting SMBH exceeds the emitted mechanical energy; 
but radiative feedback couples to the ambient medium far less 
effectively. So there is a trade-off in the effectiveness of the two 
types of feedback that must be investigated in detail.

In this paper, we extend the analysis of the purely mechanical feedback models described 
in \citet[][hereafter Paper I]{paper1}, by comparing 
the results of simulations with different feedback efficiencies against 
four well-constrained observational properties of local elliptical galaxies and their SMBHs. 
In Paper I, new hydrodynamical evolutionary 
models, combining radiative and mechanical feedback, have been developed from 
the previous purely radiative models \citep{ciotti97,
ciotti01,ciotti07}. In particular, it has been 
found that purely mechanical feedback models may not be the complete description of AGN feedback 
in Paper I. Though both feedback modes can be important together, firstly we ask in more detail 
whether popularly used purely mechanical feedback models can be accepted to explain important 
properties of SMBHs and their host galaxies in more detail. In practice, 
(1) as most commonly tested in previous research, we estimate the evolution of the ratio 
between the SMBH mass and the host galaxy stellar mass. 
This ratio has been a key diagnosis to test the formation theory of bulge-dominated galaxies 
since the correlation between two masses is found to be tight in observations. 
(2) the lifetime of 
the luminous quasar phase in the models is also compared to the observationally estimated 
lifetime (or duty cycle)\footnote{The duty cycle is commonly defined as 
the fraction of time on AGN phase 
with luminosity weights \citep[e.g.][]{ciotti01}. But the net lifetime in this paper is 
simply defined by the net duration of AGN with luminosity above some limit. 
Therefore, the net lifetime used in this paper can be slightly different from the luminosity-weighted 
AGN duty cycle multiplied by the Hubble time.} 
\citep[see][for a discussion]{martini04}. 
These two constraints are mainly governed by a small-scale physics around SMBHs. 
(3) another useful probe of the growing stellar mass in elliptical galaxies is 
the amount of the recently formed stellar mass, which is 
well constrained by observations in the rest-frame UV \citep[see][for a discussion]{yi07,kaviraj07}.
(4) we finally compare the X-ray thermal luminosity of diffuse hot gas in the models with 
the luminosity observed for local elliptical galaxies 
\citep[see][for a discussion]{pope09}. 
If the assumed AGN feedback produces 
effects which are not consistent with these observational tests, 
we may have to reconsider the adopted 
AGN feedback models and their effects in galaxy evolution.

We investigate how the four observational tests can be used to constrain the AGN feedback models, and 
how the self-regulation process works depending on the adopted models. 
If there is an appropriate efficiency of 
the mechanical AGN feedback and it is the only possible mode of feedback, then 
the purely mechanical feedback model must be consistent with 
the four observational constraints. In other words, if any values of the feedback efficiency 
cannot produce acceptable simulation results, we may need to consider the more general possibility 
which includes both radiative feedback and mechanical feedback modes, as hinted in 
Paper I. We also suggest various observable diagnostics 
which are related to the physics of AGN feedback, and which can be used to narrow 
down the feedback models and their efficiency, in addition to the four 
main tests listed above.

This paper is organized as follows. In \S2 we summarize the 
main components of two purely mechanical 
feedback models adopted for the simulations. 
In \S3 and \S4 we present the general properties of 
simulation results, and comparisons between models and observational constraints are 
given for a broad range of feedback efficiencies. Other possible tests are 
suggested in \S5, while the main results are summarized and discussed in \S6.

\section{Models}

\subsection{Simulation setup}

Here we summarize the main properties of the purely mechanical feedback models, which 
are described in detail in Paper I. In the Appendix, we summarize the important input 
physics such as mass losses from evolving stars, the various time scales of the problem, and 
the treatment of the purely mechanical and purely radiative feedback; in the present section only 
the formulae required for the discussion are given. 

In accordance to several observational constraints, at the beginning of the simulation, the galaxy model 
is described by a spherically symmetric Jaffe model for stellar mass \citep{jaffe83}, which is 
immersed in a dark matter halo, so that the total mass density follows a $r^{-2}$ profile 
\citep{ciotti09}. The effective radius of the stellar mass distribution $R_{\rm e}$ 
is about 6.91 kpc, and the line-of-sight central velocity dispersion is 260 km/s. The 
total stellar mass $M_{*}$ is $2.87\times10^{11}\ M_{\odot}$, and inside the half-mass stellar 
radius the stellar-to-dark matter mass ratio is fixed to unity. 
The obtained galaxy model obeys the 
observed Faber-Jackson relation and the Fundamental Plane of local ellipticals. 
In all models, the initial galaxy model has the central SMBH with the mass $M_{\rm BH} ~=~ 0.001 M_{*}$, 
approximately following the \citet{magorrian98} relationship. This assumption is further 
discussed in \S6.

We simulate the evolution of gas by solving 
Eulerian equations of hydrodynamics with appropriate mass, momentum, and energy sources (see Appendix). 
All calculations are conducted with a one-dimensional logarithmic radial grid 
extending from 2.5 pc to 200 kpc which has 120 bins. The simulation begins at 2 Gyr, 
i.e. a redshift of $z \sim$ 3.2 for the LCDM cosmology with 
$\Omega_{\rm m} = 0.3$, $\Omega_{\lambda} = 0.7$, and $H_{0} = 70$ km/s/Mpc, and 
continues until 14 Gyr. We note that the bulk of the mass of elliptical galaxies 
is already in place at our starting epoch. A variety of heating and cooling processes 
are included as well as spherically symmetric radiative transfer 
in several bands treated in the Eddington approximation.

We remark that, as in the previous paper \citep{ciotti01,ciotti07} and Paper I, the galaxy 
model is initially devoid of gas, and has an outflow boundary condition for the last 
radial grid point. Consequently, the ISM is solely provided by the 
recycled gas produced via stellar evolution, and the 
confining effect of the intergalactic medium is not considered. For these reasons, the obtained 
X-ray luminosity should be considered as a lower limit of the observed values \citep[see][]{pellegrini09}.

We simulate an isolated elliptical galaxy where the evolution of gas 
is not affected by any cosmological effects such as galaxy mergers and satellite 
accretions \citep{hopkins08} or just cold gas accretion \citep{khalatyan08}. 
The probability of mergers for our tested cases 
since 2 Gyr may be so small for the considered mass range 
that a passive evolution of the simulated elliptical 
galaxy is still valid \citep{maller06,kang07,drory08}. Moreover, observations of 
nearby active merger remnants show that the local physical process around the central 
SMBH such as star formation on the nuclear disk and supernovae is more important than 
what is simply described in galaxy merger simulations with AGN feedback 
\citep[e.g.][for NGC 6240]{tecza00,max07}, supporting the importance of 
our approach to include AGN feedback physics in detail. We also emphasize that 
the moderate-luminosity AGN activity at low redshift may not be caused 
by mergers even though a bright 
quasar phase may need mergers as triggers \citep{grogin05,li08}.

\subsection{Physics of purely mechanical feedback models}

As well known, there are four channels to change the total mass of gas in 
a galaxy \footnote{Because our simulation does not consider galaxy mergers and (gas) accretion, 
a possible fifth channel of gas inflow is not included in the models as we 
already mentioned in the introduction}. 
First, winds from stars leaving the main sequence, giant stars, 
planetary nebulae, SNe, etc. contribute a significant 
amount of mass in gaseous form as the galaxy evolves: over the Hubble time, 
10\% to 30\% of the initial stellar mass is ejected by stars. 
Second, when the gas cools (for spherical models usually in a cold shell), 
star formation can occur, and then the newly formed stars 
will also contribute back gas with their mass losses. 
The third channel is accretion onto the central SMBH and 
the nuclear disk \citep{shlosman90,nulsen00}. 
In fact, cooled gas can form a circumnuclear disk 
around the SMBH \citep{tan08}, and then it can be used 
to form stars in circumnuclear star bursts. 
The fate of the remaining gas in the accretion disk 
is to produce radiation 
and finally being fed into the SMBH \citep{davies07}. 
In our simulations, we also model the secondary mass loss from circumnuclear stars, 
which also formed from recycled gas (i.e. mass loss and SNe) in our models, 
that contribute to the SMBH accretion \citep{shull83,padovani93,quataert04}. 
Finally, the galactic gas can be blown out from the galaxy as a galactic wind/outflow.

Of direct interest for mechanical feedback models are the nuclear winds produced by 
the combined effects from the AGN energy output and the Type-II supernovae exploding 
in the star-forming circumnuclear disk. 
As assumed in Paper I, cold disks will form stars, and gas from the disks 
can be blown out by a 
strong wind from the central SMBH \citep{crenshaw03,krongold08}. 
In our prescription of the AGN wind loss, the 
mass loss efficiency of the AGN wind $\eta_{w}$ is defined 
to be the ratio of the disk outflow rate to the SMBH accretion rate. 
In Paper I, we introduced two different prescriptions: 
\begin{equation}
\etaw = \cases{2,\quad\quad\quad\quad\quad\quad\quad\quad\,\; {\rm 
    [A]}\cr \displaystyle{{3\etawM\over 4}{l\over 1+0.25 l}}, 
  \quad\quad\quad {\rm [B].}}
\label{eq:etaw}
\end{equation}
where $l = L_{\rm BH}/L_{\rm Edd}$ is the bolometric luminosity $L_{\rm BH}$ 
of the central SMBH in Eddington units, and the maximum wind efficiency $\eta_{w}^{M}$ 
is related to the mechanical feedback efficiency (see Equation \ref{eq:eps}). 
Therefore, as $l$ increases up to 2 (see Appendix), $\eta_{w}$ approaches $\eta_{w}^{M}$ 
in the model of type B.

In the purely mechanical feedback models, we include only the physics of mass, momentum, and energy 
transfer from the nuclear wind to the ISM \citep[see][for the description of the 
probable structure of a quasar]{elvis00}, while 
the effect of a jet is not included. 
In particular, the mechanical energy output of the circumnuclear disk is modeled as 
\begin{equation}
L_{\rm dw} ~=~ \epsilon_{\rm w} \dot{M}_{\rm BH} c^{2} ~+~ 
\epsilon_{\rm II} c^{2} (1 - f_{\rm rem,h}) \frac{M_{\rm dh*}}{\tau_{\rm *h}}, 
\label{L_dw}
\end{equation}
where 
$\dot{M}_{\rm BH}$ and $M_{\rm dh*}/\tau_{\rm *h}$ are the growth rate of the SMBH mass and 
the SFR of massive stars in the disk, respectively (see Appendix). 
In the second term of Equation \ref{L_dw}, $\epsilon_{\rm II}$ and $f_{\rm rem,h}$ are 
the feedback efficiency by Type II supernovae and the mass fraction of stellar remnants for high-mass 
stars, respectively. Therefore, when the SFR on the circumnuclear disk is high or the accretion rate to the 
SMBH is high, the total energy output from the disk wind is high with the 
contribution of the nuclear wind modulated by $\epsilon_{\rm w}$.

Our main concern in this paper is $\epsilon_{\rm w}$ which determines how efficiently 
the growing central SMBH can deposit the mechanical energy into its surrounding ISM. 
Following Paper I, we test two models:
\begin{equation}
\epsw\equiv\cases{\epswM,\quad\quad\quad\quad\quad\quad\quad\;\;\, {\rm [A]}\cr
                 \displaystyle{{3\epswM\over 4}{l\over 1+0.25 l}},
\quad\quad\quad {\rm [B].}} 
\label{eq:eps}
\end{equation}
In the both descriptions, $\eta_{\rm w}^{\rm M} = 1800 \epsilon_{\rm w}^{\rm M}$ so that 
the velocity of the circumnuclear wind is about $10^{4}$ km/s in accordance with 
observations of broad-line winds \citep{crenshaw03}.

In this paper, we maintain the nomenclature of Paper I, and we refer to 
purely mechanical feedback models of type A and B as MA and MB models, respectively. 
It is finally important to recall that, even though radiative feedback effects are not 
considered in the simulations, $L_{\rm BH}$ has a fundamental importance in the MB models, 
as $\epsilon_{\rm w}$ is dependent of the luminosity.

The popular choice of 
$\epsilon_{\rm w}^{\rm M} = 0.005$ in type A models has been adopted in 
the majority of previous research. 
For example, 0.5\% of the accretion mass is immediately deposited to the ISM as a feedback 
energy, while assuming that 10\% of the accreted mass is converted to radiation \citep{soltan82} in 
simulations of galaxy mergers \citep{springel05a,springel05b,hopkins05,johansson09}. 
\citet{thacker06} also examine a case that corresponds to our model A 
with $\epsilon_{\rm w}^{\rm M} = 0.005$, which easily heats up gas within a few Mpc. 
Other examples are found in investigations of the 
Sunyaev-Zel'dovich effect from quasars \citep[e.g.][]{chatterjee08,scannapieco08}, 
downsizing of galaxy evolution \citep[e.g.][]{scannapieco05}, the evolution of the 
black hole mass and bulge mass \citep[e.g.][]{robertson06}, 
and the formation of elliptical galaxies \citep[e.g.][]{khalatyan08}. 
However, ab initio we do not know of either an empirical basis for the adoption of this 
particular coefficient nor calculations from which it could be derived. 

There have been some trials to pin down an acceptable range of $\epsilon_{\rm w}^{\rm M}$. 
For example, \citet{scannapieco04} investigate the cases 
corresponding to the MA models with $\epsilon_{\rm w}^{\rm M} = 0.0025$ and 0.01, and conclude 
that the two values fail to explain quasar luminosity function and other structure formations. But 
as they note, $\epsilon_{\rm w}^{\rm M}$ is observationally not well constrained by their 
tests. \citet{levine05} also examine how a constant mechanical feedback efficiency 
affects the distribution of AGN outflows by testing 
$\epsilon_{\rm w}^{\rm M} = 0.001$, 0.01, and 0.005. 
They find $\epsilon_{\rm w}^{\rm M} = 0.001$ as the best value by comparing 
their models with observationally derived filling fractions of 
AGN outflows. As these two examples show, 
the acceptable range of $\epsilon_{\rm w}^{\rm M}$ is still controversial 
for the MA models. However, 
the different purely mechanical models such as our MB models have not been tested 
extensively yet, even though 
they appear to be more close to the real AGN feedback process in which the 
feedback efficiency is expected to increase as one approaches the Eddington limit 
\citep{kurosawa09}.

For these reasons, we test eight different 
values of $\epsilon_{\rm w}^{\rm M}$ for both model MA and MB 
ranging from $1\times10^{-5}$ to $5\times10^{-2}$ (see Table 
\ref{tab:epsilon}). 
Because our goal is to constrain the range of acceptable efficiencies and models with the purely 
mechanical feedback, the tests also include unreasonably low and high values of 
$\epsilon_{\rm w}^{\rm M}$. We note that a model without AGN feedback effects was shown completely 
unreasonable in Paper I, producing an extremely massive SMBH.

\section{Results of the MA models}

As we explained in the previous section, the family of the MA models, 
i.e. the mechanical feedback models 
with a fixed feedback efficiency, has been widely adopted in various papers as the simplest 
prescription of AGN feedback. 
In Paper I, it is shown how these models are characterized by a quite rigid behavior with 
a sharp transition from very strong feedback effects to almost no effects, 
even though the input physics is quite elaborate. Here, we explore 
other aspects of these models, starting with the 
commonly used $\epsilon_{\rm w}^{\rm M} = 0.005$ (Run 6), and then progressively 
reducing it from Run 6 to Run 1 and increasing from Run 6 to Run 8.

\subsection{Evolution of the MA models} 

\begin{figure}[t!]
\plotone{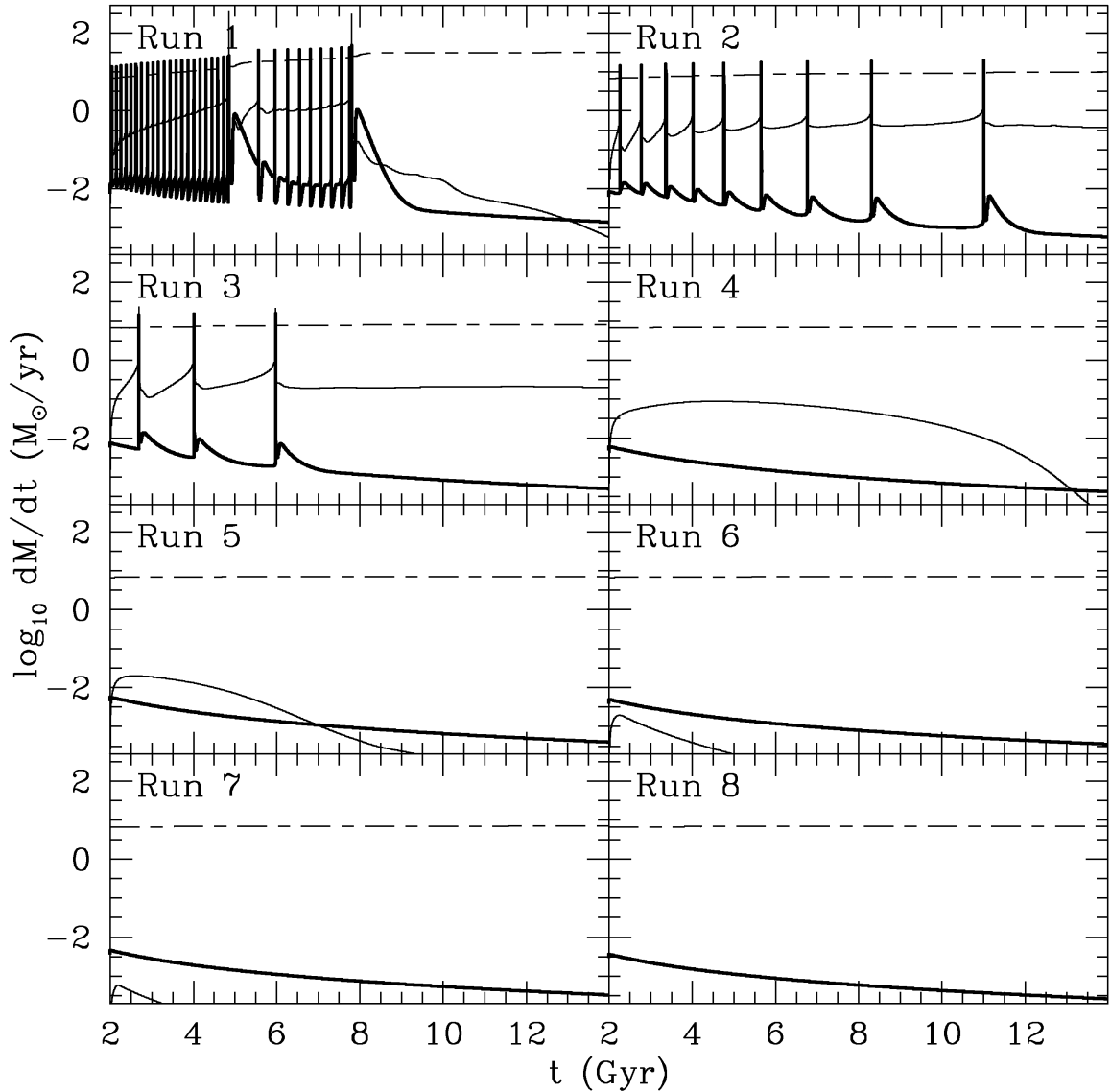}
\caption{SMBH mass accretion rate $\dot{M}_{\rm BH}$ ({\it thick line}) 
and the SFR $\dot{M}_{*}$ ({\it thin line}) in the MA models. 
Both rates show a strong time-dependence. 
The models with the lower feedback 
efficiency generally produces higher 
$\dot{M}_{\rm BH}$ and $\dot{M}_{*}$ more frequently. 
The Eddington accretion rate ({\it short-long dashed line}) 
is usually higher than $\dot{M}_{\rm BH}$ except for the peak activities.
}
\label{fig:MA_time1}
\end{figure}

\begin{figure}[t!]
\plotone{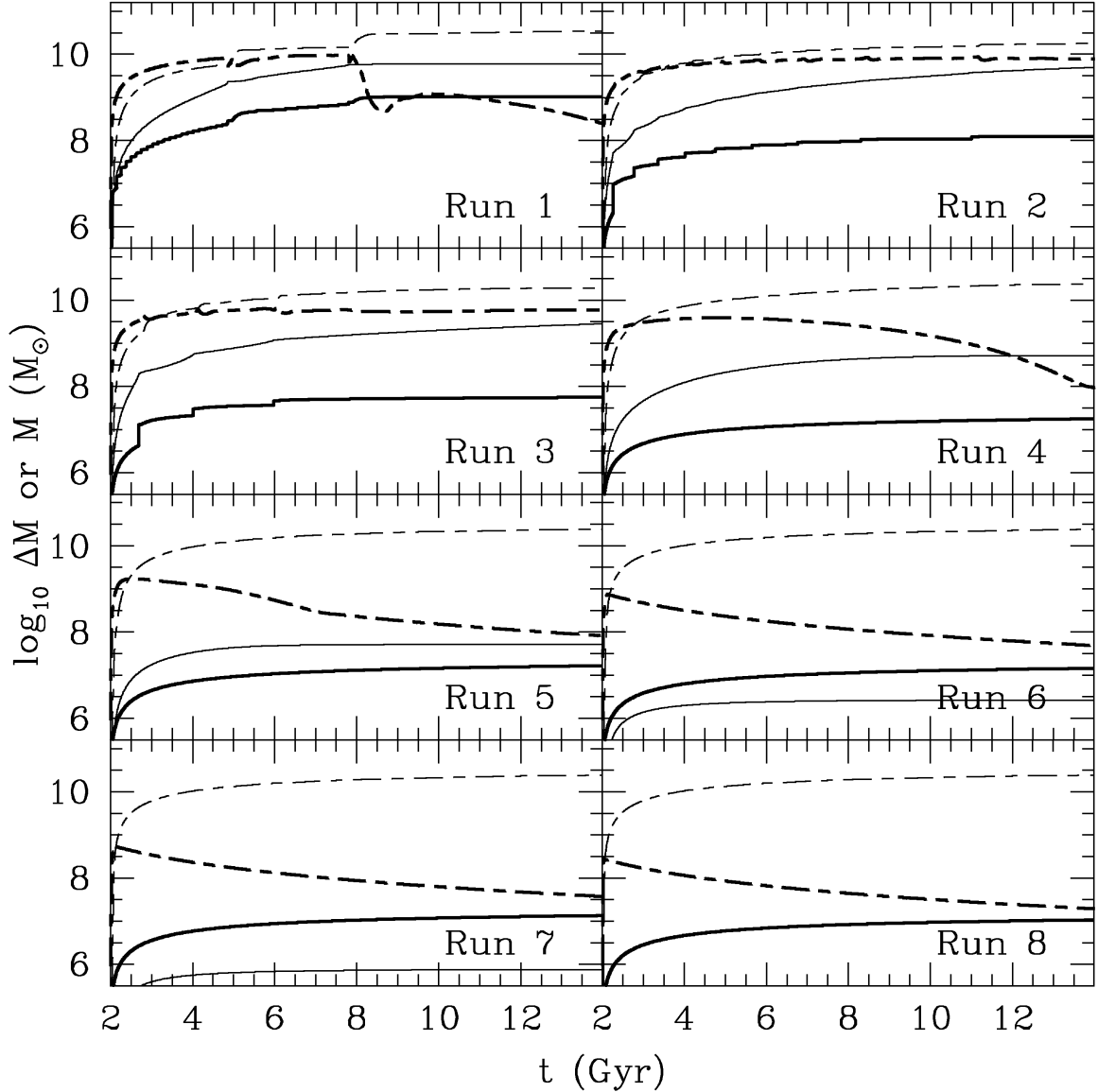}
\caption{Time evolution of the mass budget in the MA models: 
the change of the SMBH mass $\Delta {M}_{\rm BH}$ ({\it thick solid line}), 
the change of the stellar mass $\Delta {M}_{*}$ ({\it thin solid line}), 
the total mass of gas within 10 $R_{\rm e}$ in the galaxy $M_{\rm g}$ ({\it thick short-long dashed line}), 
and the time-integrated mass of blown-out gas $\Delta {M}_{\rm w}$ ({\it thin short-long dashed line}). 
Following the difference in $\dot{M}_{\rm BH}$ and 
$\dot{M}_{*}$, models with different efficiencies result in different evolution 
of the mass ratio between the central SMBH and stars. $M_{\rm g}$ decreases because 
the total SMBH accretion rate, star formation rate, and galactic wind rate dominate at late times 
over the stellar mass loss rate.
}
\label{fig:MA_time2}
\end{figure}

Figures \ref{fig:MA_time1} and \ref{fig:MA_time2} present the time evolution of 
some global quantities in the MA models. 
In all runs, the peak of the SFR ($\dot{M}_{*}$) is reached 
earlier than the peak of the SMBH mass accretion rate ($\dot{M}_{\rm BH}$); 
in some cases the two 
peaks are coincident within the limit of the simulation time resolution. 
High feedback efficiencies generally delay an initial star-burst and a high 
$\dot{M}_{\rm BH}$, and stop early further star formation and SMBH mass accretion. 
As found in the simulations from the Run 4 to 8 
with high values of $\epswM$, the effective time-scale of 
the feedback process is so short 
that any peaked formation of stars and 
mass accretion to the central SMBH does not occur at all. 
Meanwhile, the low feedback efficiencies make galaxies have 
extended star formation and SMBH mass accretion with multiple peaks. 

Remarkably, 
the main difference caused by different feedback efficiencies is discovered in the rise in 
stellar mass $\Delta M_{*}$ and SMBH mass $\Delta M_{\rm BH}$. If the initial 
ratio $M_{\rm BH}/M_{*}$ needs to be maintained at all times, 
$\Delta M_{\rm BH}/\Delta M_{*}$ is required to follow the same initial ratio $\sim 10^{-3}$. 
Yet, $\Delta M_{*}$ is always higher than $\Delta M_{\rm BH}$ 
for $\epsilon_{\rm w}^{\rm M} = 5 \times 10^{-4}$, 
while this trend is reversed for high feedback efficiencies, as shown in Figure \ref{fig:MA_time2}. 
Therefore, our simulations imply extreme fine-tuning of the efficiency $\epsilon_{\rm w}^{\rm M}$ 
in order to maintain the SMBH to stellar mass ratio at a constant level.

\begin{figure}[t!]
\plotone{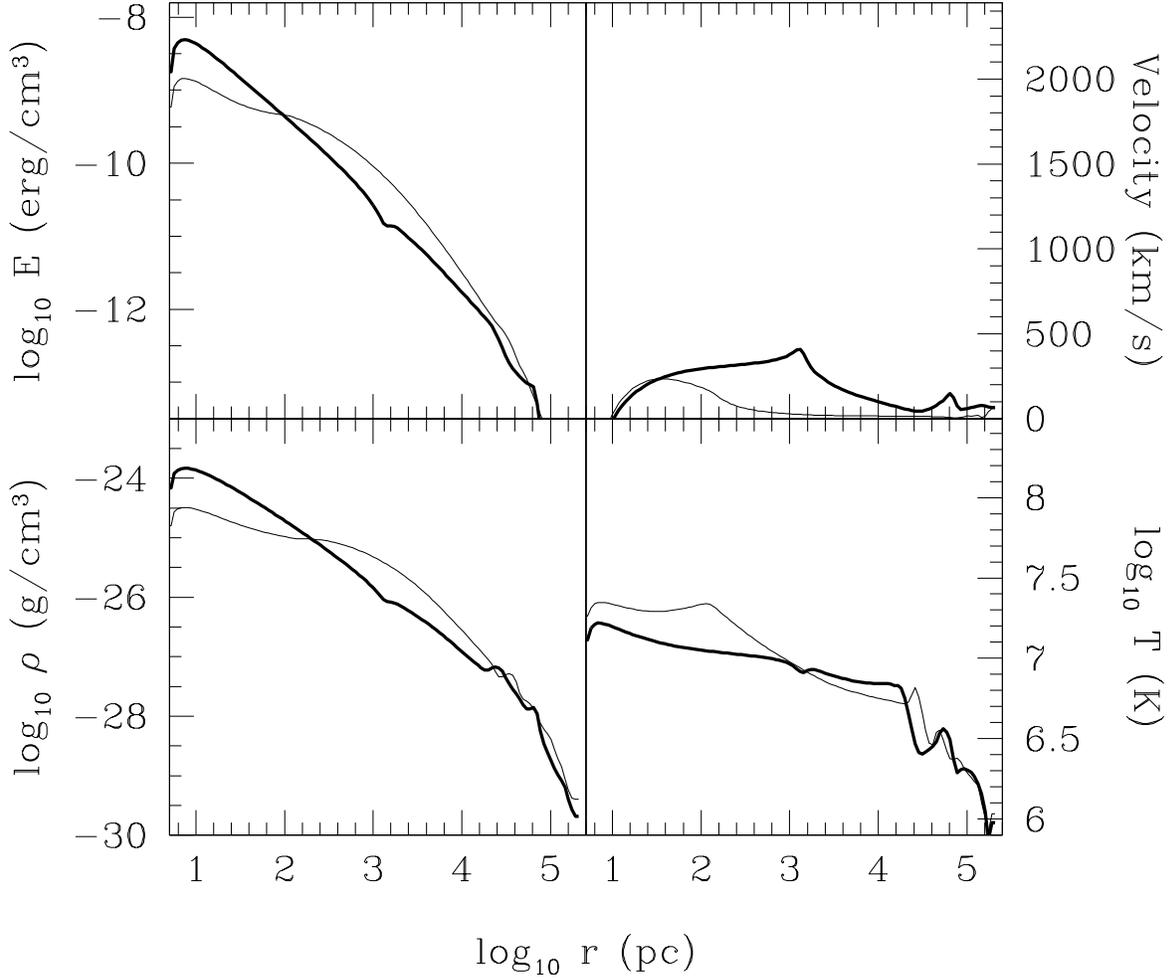}
\caption{Radial structure of the gas 
internal energy density, velocity, mass density, and temperature 
in the MA model with 
$\epsilon_{\rm w}^{\rm M} = 5 \times 10^{-5}$ (Run 2), 
at t = 3.5 Gyr ({\it thick line}) and 5.5 Gyr ({\it thin line}), 
corresponding to inter-burst phases (see Figure \ref{fig:MA_time1}). 
The distributions of internal energy density and mass density 
do not change significantly between the two epochs.
}
\label{fig:MA_space1}
\end{figure}

\begin{figure}[t!]
\plotone{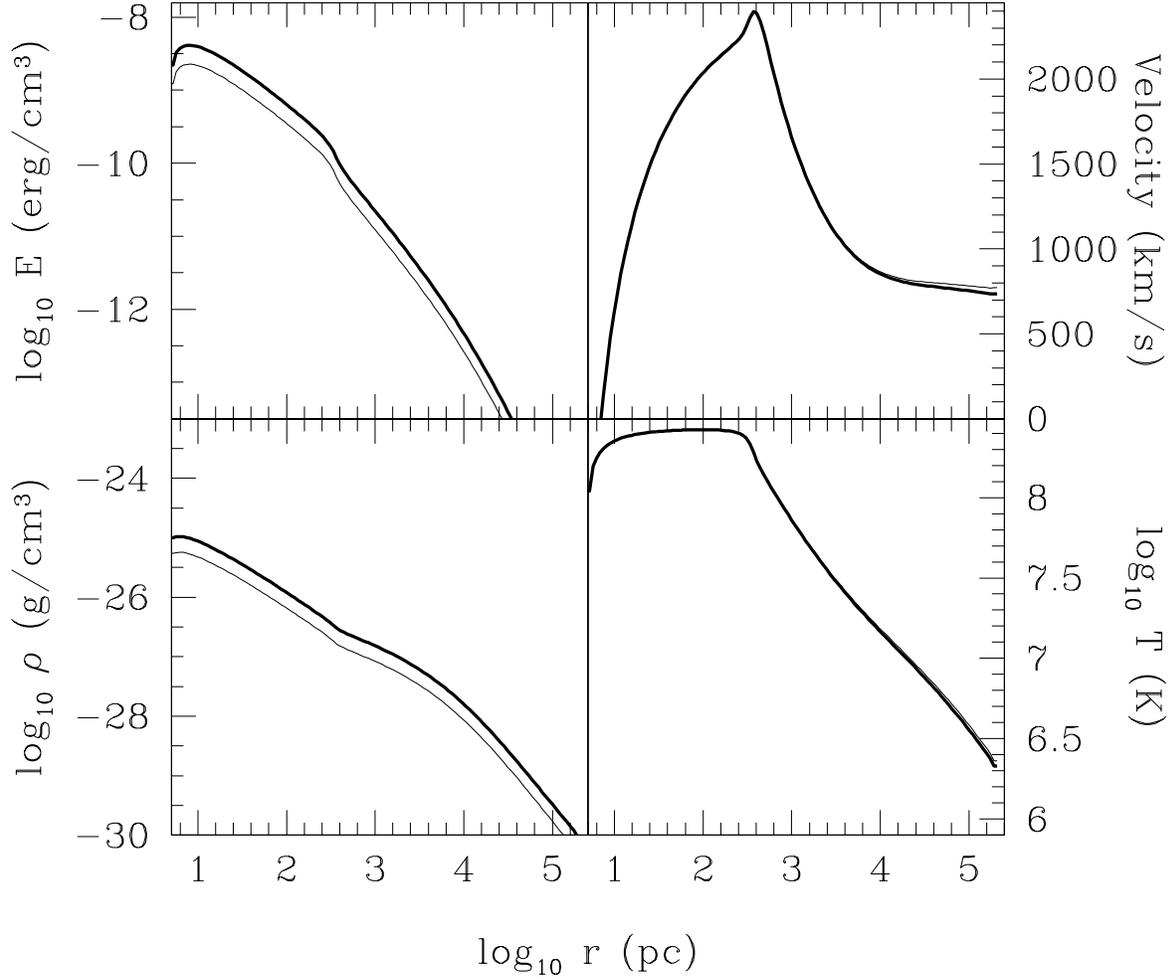}
\caption{Radial structure of gas properties in the MA model 
with $\epsilon_{\rm w}^{\rm M} = 5 \times 10^{-3}$ (Run 6) 
at the same epochs as Figure \ref{fig:MA_space1}. 
The comparison with Run 2 (Figure \ref{fig:MA_space1}) 
shows that the velocity field and temperature distribution 
strongly depend on the feedback efficiencies. The high 
feedback efficiency induces the high-velocity outflow, 
the increase in gas temperature which is characterized by 
a flat high-temperature region of about 100 pc, and the 
corresponding low value of the central density. 
This trend is consistent with the overall suppression of both mass accretion to the SMBH and 
star formation, as seen in Figure \ref{fig:MA_time2}.
}
\label{fig:MA_space2}
\end{figure}

Different feedback efficiencies also affect the radial structure of gas. As shown in 
Figures \ref{fig:MA_space1} and \ref{fig:MA_space2}, 
models with the high feedback efficiency are more likely to 
produce a high-velocity outflow at a large distance. For example, at $r \sim$ 20 kpc, 
the outflow velocity is about 800 km/s for Run 6 ($\epsilon_{\rm w}^{\rm M} = 5 \times 10^{-3}$). 
But around the peak time 
of the SMBH mass accretion, the outflow in Run 2 
($\epsilon_{\rm w}^{\rm M} = 5 \times 10^{-5}$) 
has much lower velocity than 500 km/s at the same distance. 
High feedback efficiencies enhance the outflow 
as pointed out by \citet{scannapieco04} and \citet{levine05}, even though SFR in 
their cases is not as high as in our simulation \citep{david06}.

Heating by stellar processes and the additional AGN heating alters the 
temperature distribution of the gas within galaxies. 
As already discussed in numerous previous research \citep[e.g.][]{tabor93,binney95,ciotti97,
ciotti01,cattaneo07}, energy deposited by mechanical feedback heats up the surrounding 
ISM, resulting in a core structure of hot gas such as seen in Figures \ref{fig:MA_space1} and 
\ref{fig:MA_space2}. Although 
the high feedback efficiency suppresses further star formation and stellar feedback processes, 
even in these cases the energy and mass supply by the strong AGN wind 
compensates for the lack of stellar feedback processes and finally heats up the gas.

\begin{figure}[t!]
\plottwo{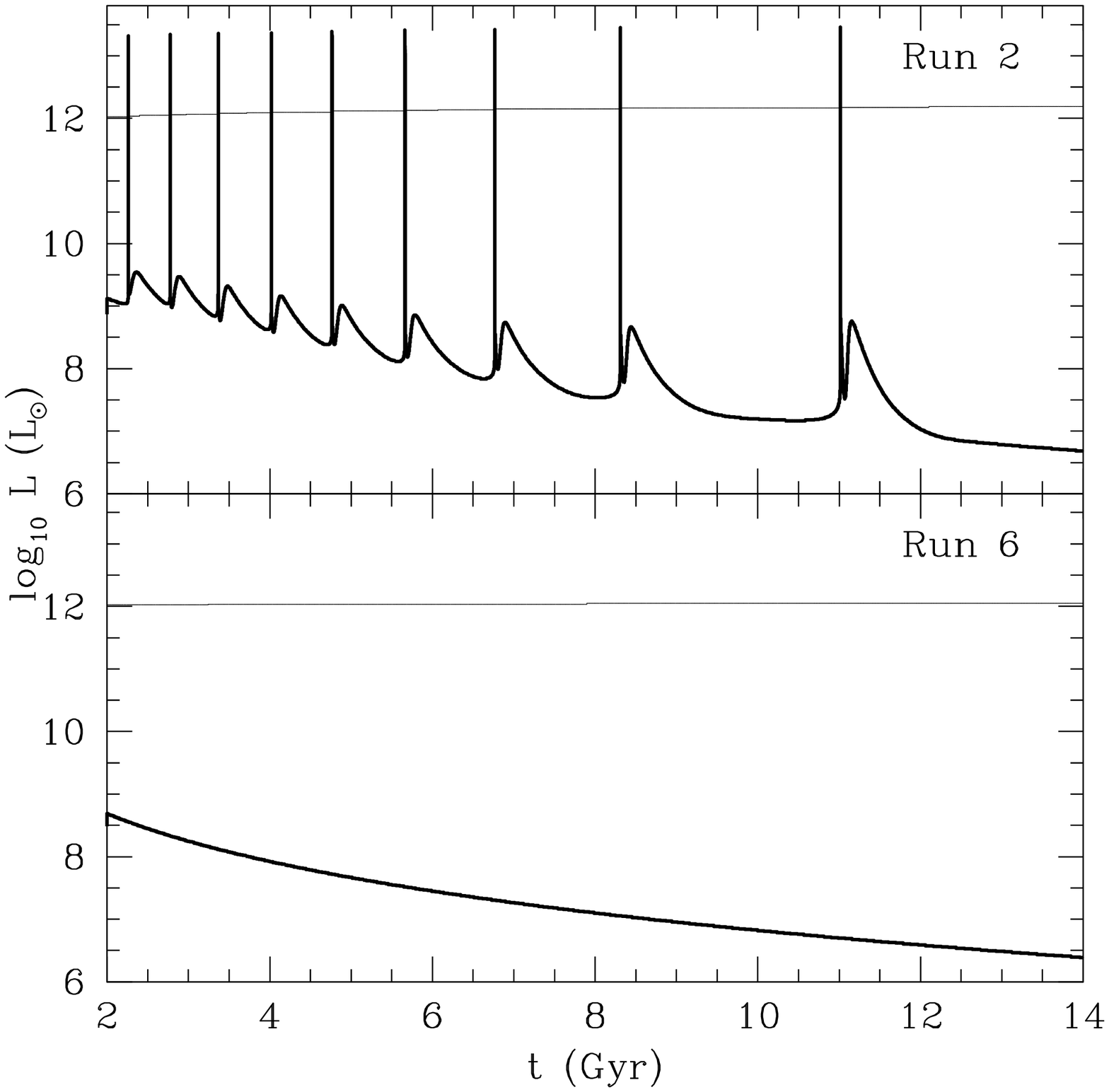}{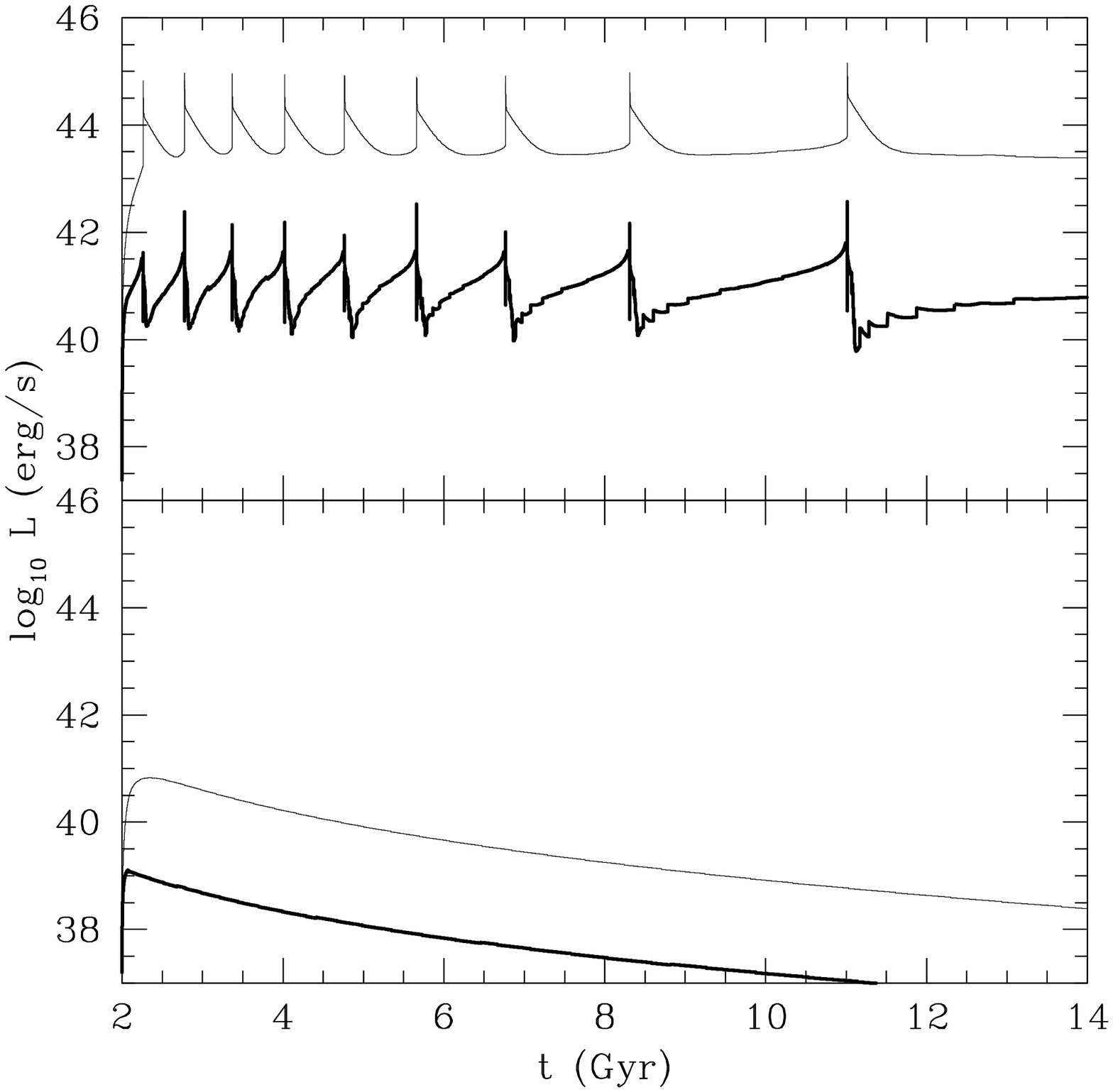}
\caption{
Time evolution of the dust-extincted optical accretion luminosity $L_{\rm opt}$ 
(left panel) and the X-ray luminosity from hot ISM within 10 $R_{\rm e}$ 
and infrared re-emission of stellar radiation (right panel) for Run 2 and 6. 
$L_{\rm opt}$ ({\it thick solid line}) is basically coupled to 
the mass accretion history shown in Figure \ref{fig:MA_time1}, but it is 
usually much lower than 
the 10\% of the Eddington luminosity ({\it thin solid line}). 
Only models with low 
feedback efficiencies achieve the optical luminosity which is higher than the 10\% of the 
Eddington luminosity for short periods. 
The overall change of the X-ray luminosity from hot gas ({\it thick line}) and the infrared 
re-radiation of stellar light by dust ({\it thin line}) 
follows the evolution of SFR rather than the history of the SMBH mass 
accretion. 
But right after the peaks of the SMBH accretion rate in Run 2, 
the X-ray luminosity ({\it thick line}) 
declines more quickly than the infrared emission. 
This difference implies that the X-ray luminosity 
may be a more sensitive probe to measure the effects of AGN feedback on 
the ISM than IR emission.
}
\label{fig:MA_lum}
\end{figure}

As shown in Figure \ref{fig:MA_lum}, the time evolution of the accretion luminosity 
is strongly coupled to the SMBH mass accretion history. 
However, we note that the luminosity at t $\sim$ 14 Gyr is similar 
even for different values of feedback efficiencies, because 
the difference in the SMBH mass accretion rate is small at late times despite the 
difference of $\epsilon_{\rm w}^{\rm M}$. $L_{\rm BH}$ 
is generally lower than the Eddington luminosity for both 
high and low feedback efficiencies. Bursts of high $L_{\rm BH}$ above the Eddington 
luminosity are found only in the models with low feedback efficiencies such as Run 2. 
As we explain in Appendix, our AGN model permits a moderate super-Eddington accretion. 
The AGN feedback with the low efficiency is not strong enough to stop the 
development of cold gas when the precursor of a large amount of cold gas reaches the central 
region. Therefore, a large fraction of the cold gas is accreted to the central SMBH at the end. 

We find that the star formation history is mainly parallel to the evolution of both the 
X-ray luminosity from the hot diffuse ISM and the infrared (IR) emission by dust, 
which re-radiates the absorbed stellar radiation, as shown in Figure \ref{fig:MA_lum}. 
As the difference in the star formation history implies 
in Figure \ref{fig:MA_time1}, the difference in X-ray and IR luminosity is more 
significant for different feedback efficiencies than the difference in $L_{\rm BH}$. 
In addition, the X-ray luminosity of hot ISM intricately changes more than 
the IR emission, 
corresponding to the energy input from the AGN feedback and the mass loss by the 
continuously escaping hot gas (see Figure \ref{fig:MA_time2}).

\subsection{Are the MA models acceptable?}

\begin{figure}[t!]
\plotone{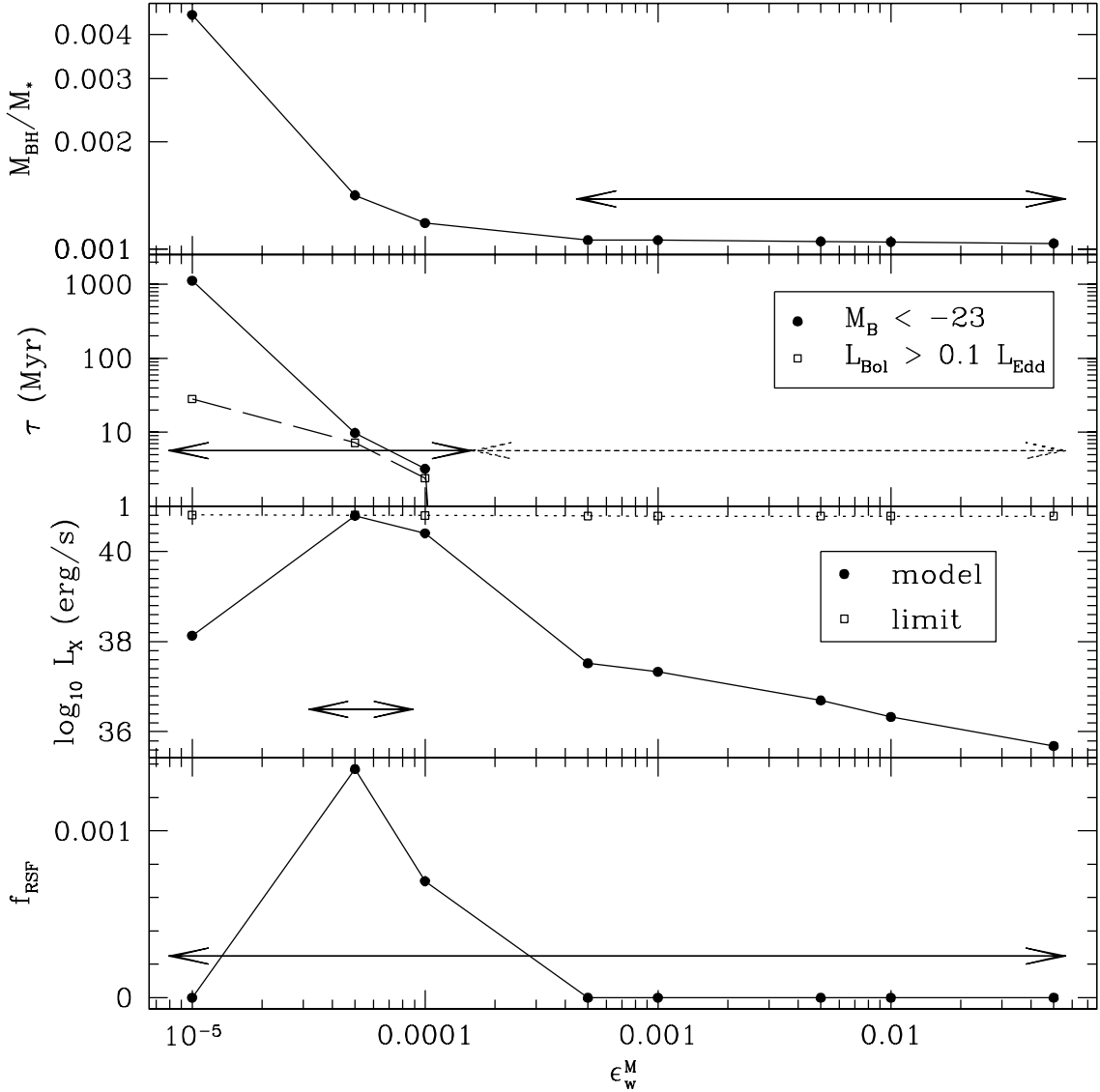}
\caption{Results of the four proposed tests for the MA models at the epoch of 14 Gyr. 
From top to bottom, 
four tests are the SMBH mass to stellar mass ratio, 
the net quasar lifetime, the X-ray luminosity of the 
diffuse hot gas inner 10 $R_{\rm e}$, and the fraction of stellar mass formed within the last 1 Gyr 
$f_{RSF}$. 
The ratio $M_{\rm BH}/M_{*}$ in simulations can be consistent with 
observations for $\epsilon_{\rm w}^{\rm M} 
> 1 \times 10^{-4}$. Yet, $L_{X}$ is inconsistent for $\epsilon_{\rm w}^{\rm M} 
> 1 \times 10^{-4}$, and the corresponding models do not experience quasar-like phases. The arrow 
bars represent the range of $\epsilon_{\rm w}^{\rm M}$ that is acceptable for each test. For the net 
quasar lifetime, $\epsilon_{\rm w}^{\rm M} \gg 10^{-4}$, which is represented by the dotted arrow bar, 
is acceptable only if our models do not 
need to experience any quasar phases in the evolution.
}
\label{fig:MA_comp}
\end{figure}

Figure \ref{fig:MA_comp} summarizes the results of our tests on the MA models. 
The mass ratio between the central SMBH and the host galaxy at 14 Gyr, i.e. $z \sim$ 0, 
monotonically increases as the feedback efficiency 
decreases. For $\epsilon_{\rm w}^{\rm M} > 1 \times 10^{-4}$, 
the growth of stellar and SMBH mass is so small 
that the deviation from the initial ratio is not significant. Following the increased 
SMBH accretion, the net lifetime of bright AGN phase is 
also long in the models with the low $\epsilon_{\rm w}^{\rm M}$. When we adopt 
the B-band magnitude $M_{\rm B} < -23$ mag as a limit of optical quasars \citep{martini04} 
and use the typical spectral energy distribution of quasars \citep{elvis94}, the 
observational constraint on the maximal net lifetime is about 1 Gyr \citep{martini04}. 
Even though there is no well defined observational limit 
on the net lifetime when the bolometric luminosity $L_{\rm BH}$ 
is higher than 10\% of the Eddington luminosity $L_{\rm Edd}$, \citet{hopkins05} claims that it is 
comparable to the lifetime of the optical limit on quasars. We find that the two measurements 
can be comparable for only $\epsilon_{\rm w}^{\rm M} \gg 10^{-5}$. But 
if our models have to reproduce any quasar phases since 2 Gyr, 
$\epsilon_{\rm w}^{\rm M} \gg 10^{-4}$ is not acceptable because it does not permit any 
strong SMBH accretion phases as shown in Figure \ref{fig:MA_time1} and does not experience 
any luminous phases with $M_{\rm B} < -23$ mag or $L_{\rm BH} > 0.1 L_{\rm Edd}$.

\begin{figure}[t!]
\plotone{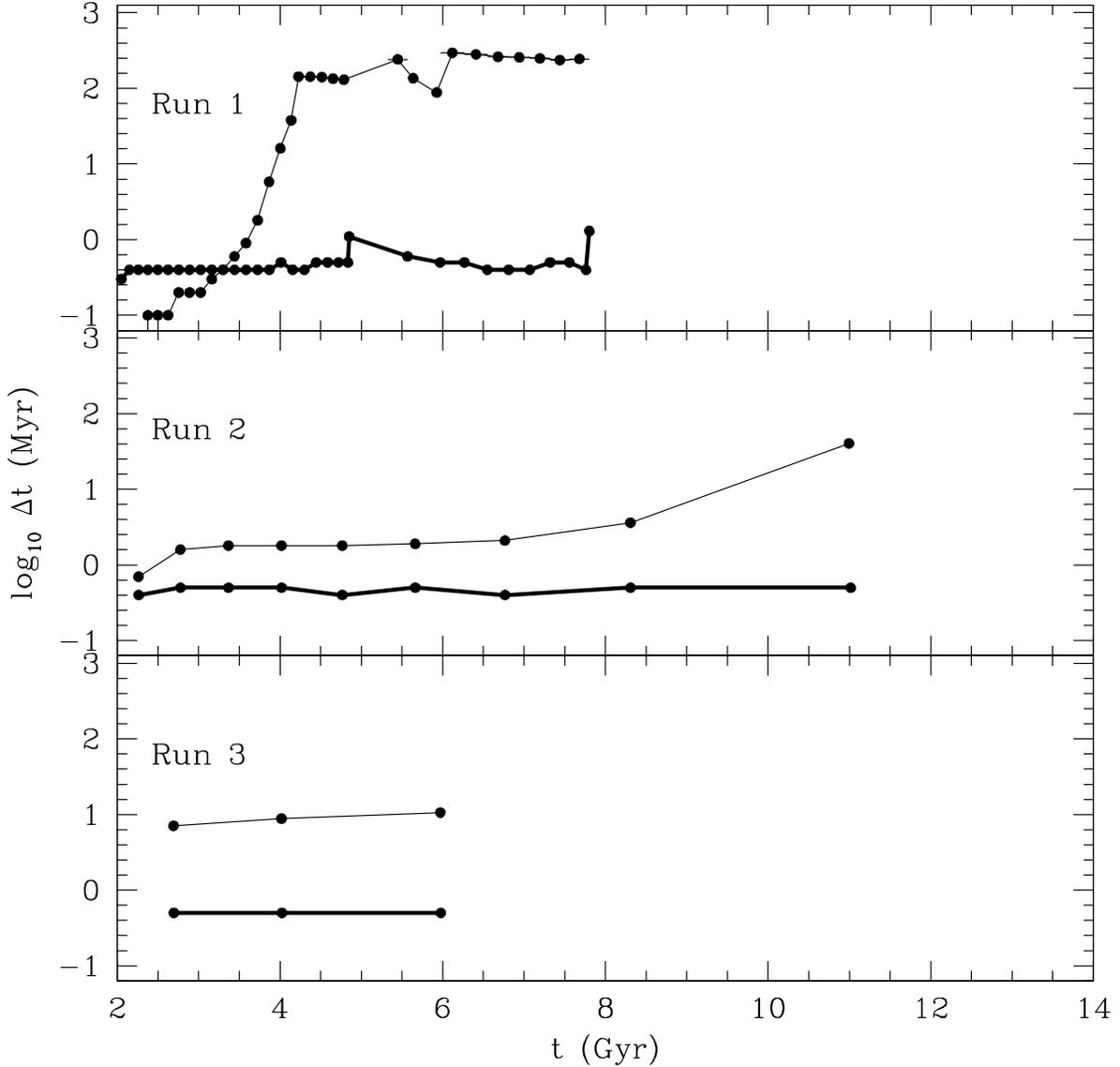}
\caption{Episodic lifetime of intensive star formation and quasar activity in the MA models. For 
Run 1, 2, and 3, we measure 
the episodic lifetime of SFR $\dot{M}_{\odot} >$ 1 $M_{\odot}/yr$ ({\it thin line}) and when 
the bolometric luminosity from the central SMBH $L_{\rm BH}$ is higher than the Eddington luminosity 
$L_{\rm Edd}$ ({\it thick line}). The typical episodic lifetime 
of $L_{\rm BH} > L_{\rm Edd}$ is about 0.4 
Myr for all three simulations at any time. The high SFR generally shows a long 
duration at late time.}
\label{fig:MA_dT}
\end{figure}

The episodic quasar lifetime can be used to constrain our models in addition to the net 
quasar lifetime. The episodic lifetime is measured for each instance of the quasar phase, i.e. 
the high SMBH accretion phase, while the net lifetime is the sum of the episodic lifetime for all 
instances. Because our models do not provide any information about quasar 
activity before the initial time of simulations, i.e. 2 Gyr, the net quasar lifetime of 
the models can be increased up to 2 Gyr further. Therefore, the 
episodic quasar lifetime 
can be a better diagnostic in our simulations. In Figure \ref{fig:MA_dT}, we present the 
change of the episodic lifetime for $L_{\rm BH} > L_{\rm Edd}$ as an example. 
This episodic lifetime does not change significantly for every episodic activity, having a 
typical duration 0.4 Myr which is longer than both a fixed simulation output time-step size 0.1 Myr and 
varying computational time-step sizes (in average, about 50 year). The episodic lifetime for $M_{\rm B} < -23$ mag also has 
the same pattern, but is about 1 Myr which is acceptable compared to a recent measurement 
\citep{kirkman08}. 

We also compare the predicted X-ray luminosity of the hot ISM and 
the mass fraction of recently formed stars 
to the available observational constraints. The constraint on the X-ray luminosity is 
derived from the typical X-ray luminosity of local ellipticals \citep{osullivan01} after subtracting 
the contribution of discrete X-ray sources from the X-ray luminosity \citep{ciotti91,kim04}. 
As shown in Figure \ref{fig:MA_comp}, if the 
feedback efficiency is too low as in the model with $\epsilon_{\rm w}^{\rm M} = 1 \times 10^{-5}$, 
too much mass 
is accreted to the central SMBH, decreasing the X-ray luminosity and suppressing 
the late star formation. Meanwhile, the low SFR in the models with high feedback 
efficiencies causes the lack of hot gas, which sequentially results in low X-ray luminosities. 
Although the temperature of the hot gas is higher in models with the 
high feedback efficiencies, the total amount of gas is too low to be compensated by the high 
temperature. 
However, it is important to recall that all models explored in this paper represent a galaxy 
that is initially devoid of gas and without external pressure \citep[see][]{pellegrini09}. 
The measured recently formed stellar mass for the last Gyr is lower than 
a few percent in local elliptical galaxies \citep{yi07,donas07,kaviraj07}, 
which is much higher than what we find in our MA models. Hence, the fraction of recently formed 
stellar mass in simulation results is acceptable compared to local ellipticals.

The duration of SFR $M_{*} > 1 M_{\odot}/yr$ depends on when intensive star formation 
occurs and how strong the feedback efficiency is. 
Generally, the late star formation has a longer duration. For example, the late 
star formation in Run 1 continues longer than 100 Myr, as shown in Figure \ref{fig:MA_dT}. 
The effect of different feedback efficiencies is significant in modulating the early star formation. 
For example, in Run 3 the peaked early star formation lasts for about 10 Myr. Yet, the low 
feedback efficiency in Run 1 allows the accretion to the central SMBH 
to occur more frequently, resulting in 
the short duration of vigorous star formation.

In short, it is difficult to find the range of the feedback efficiency in the MA model 
that satisfies the four observational constraints together. Low efficiencies 
($\epsilon_{\rm w}^{\rm M} < 5\times10^{-4}$) produce too 
massive central SMBHs or long net lifetime of quasar activity. High efficiencies 
($\epsilon_{\rm w}^{\rm M} > 5\times10^{-5}$) have 
different problems: too low X-ray luminosity or no quasar activity. 
Importantly, the popularly 
used $\epsilon_{\rm w}^{\rm M} = 0.005$ does not pass the four tests simultaneously. 

\section{Results of the MB models}

We now move to discuss the MB models. In these models, the mechanical output 
from the central SMBH depends on the accretion luminosity, 
increasing as a function of $L_{\rm BH}/L_{\rm Edd}$ as presented 
in Equation \ref{eq:eps}. 
This description is definitely more 
close to the real processes around the central SMBHs than 
the fixed mechanical feedback efficiency in the MA models 
\citep{kurosawa09}. 
As we will see, however, 
the basic consequence of varying the feedback efficiency is 
qualitatively same in MA and MB models, despite their differences.

\subsection{Evolution of the MB models}

\begin{figure}[t!]
\plotone{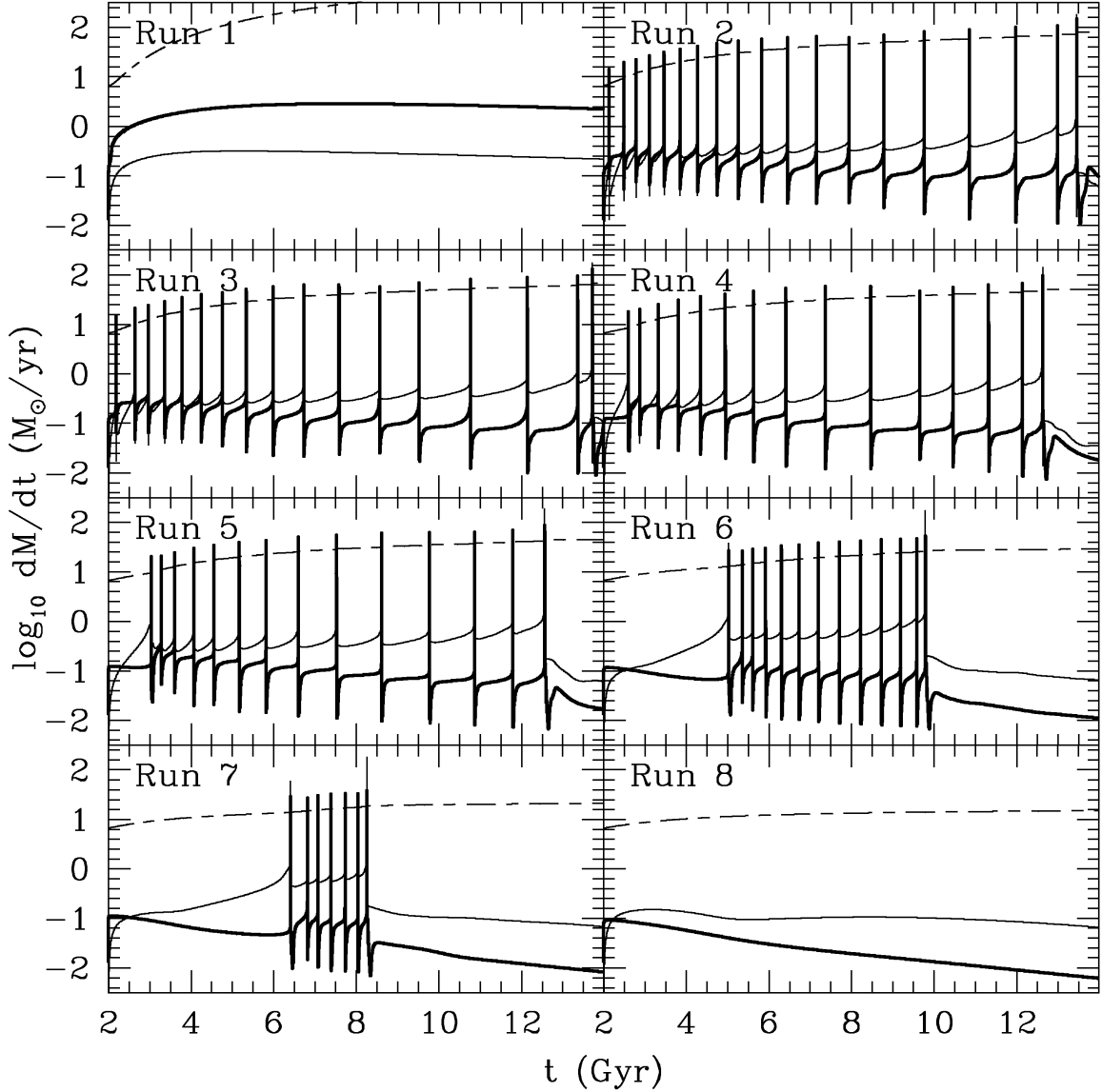}
\caption{
SMBH mass accretion rate $\dot{M}_{\rm BH}$ ({\it thick line}) 
and SFR $\dot{M}_{*}$ ({\it thin line}) in the MB models. 
The Eddington accretion rate ({\it short-long dashed line}) 
is always much higher than $\dot{M}_{\rm BH}$ except for the peak activities, 
as found in the MA models. 
In the models with either extremely high or low feedback efficiencies, 
self-regulated outbursts are not found, showing almost constant or decreasing 
$\dot{M}_{\rm BH}$ and $\dot{M}_{*}$. Compared with the MA models, the MB models substantially 
increase the number of bursts because $\epsilon_{\rm w}$ is low at low accretion luminosity, 
and this favors a quick accumulation of recycled gas in the galaxy.
}
\label{fig:MB_time1}
\end{figure}

The peak value of the feedback efficiency $\epsilon_{\rm w}^{\rm M}$ 
determines how frequently the SMBH and its host galaxy can 
achieve high SMBH mass accretion rate and SFR, as we already found in the MA models. Figure 
\ref{fig:MB_time1} presents the SMBH mass accretion rate and SFR in the MB models. 
Either extremely high or low feedback efficiencies does not permit the 
resurrection of both high $\dot{M}_{\rm BH}$ and $\dot{M}_{*}$. 
For example, in Run 1 SFR is 
higher than the SMBH mass accretion rate at any time, while the reversed pattern 
of growth rates is found in Run 8.

The peak feedback efficiency also determines the onset of the earliest burst and the last 
burst in $\dot{M}_{\rm BH}$ and $\dot{M}_{*}$. 
In Run 1 with $\epsilon_{\rm w}^{\rm M} = 1 \times 10^{-5}$, 
the initial effect of mechanical feedback is too weak 
to suppress the growth of the SMBH mass, even though star formation is always calm. 
As seen from Run 2 to 7 in Figure \ref{fig:MB_time1}, 
increasing $\epsilon_{\rm w}^{\rm M}$ prevents the early intensive mass accretion onto the BH, 
but the global SFR is not initially affected by the high feedback efficiencies. 
High feedback efficiencies 
also cause the early cessation of repeating peaked high $\dot{M}_{\rm BH}$ and $\dot{M}_{*}$. 
For example, the last peak $\dot{M}_{\rm BH}$ of Run 7 is found to be about 5 Gyr earlier than 
that of Run 5.

\begin{figure}[t!]
\plotone{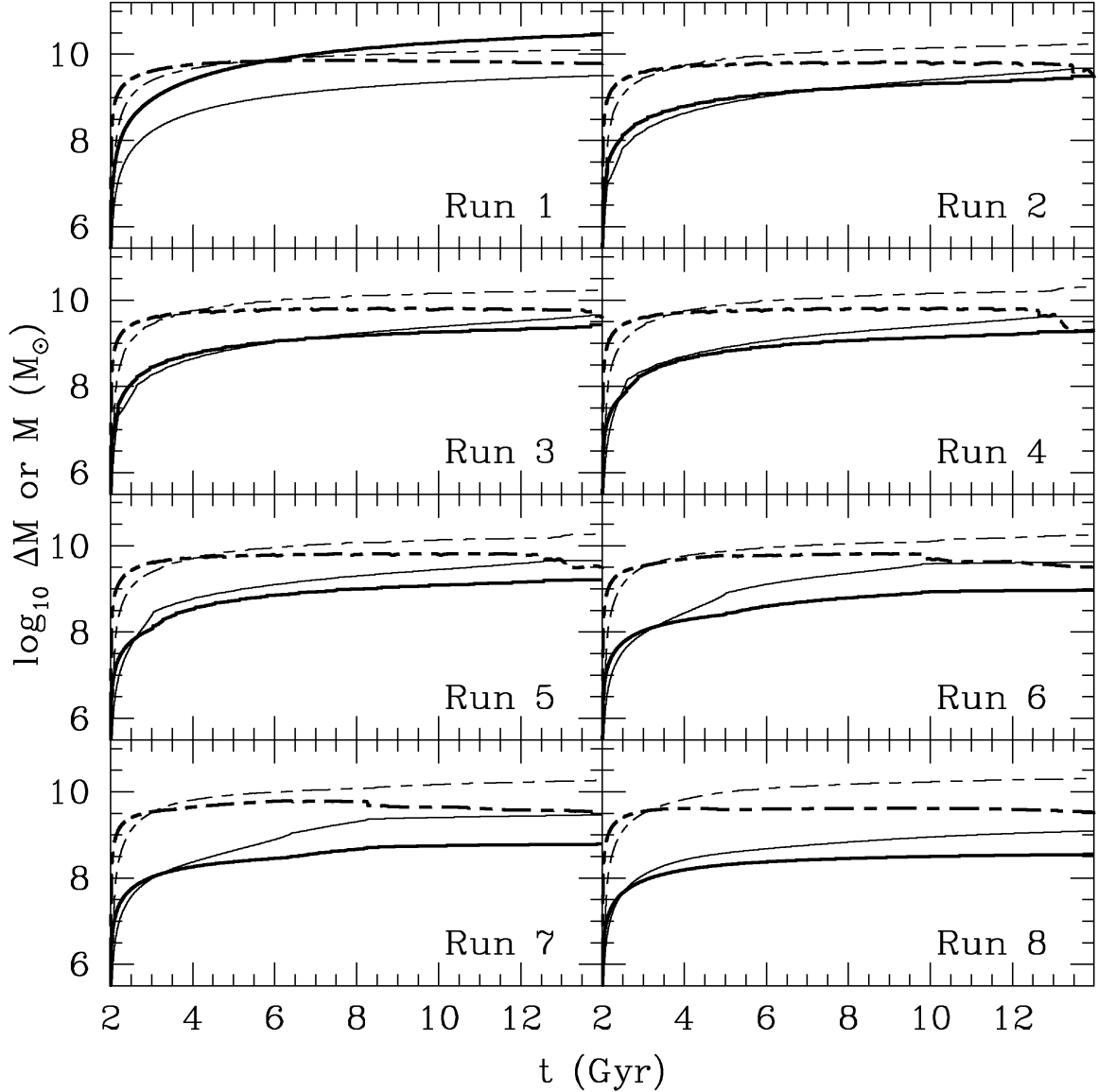}
\caption{
Time evolution of the mass budget in the MB models: 
the change of the SMBH mass $\Delta {M}_{\rm BH}$ ({\it thick solid line}), 
the change of the stellar mass $\Delta {M}_{*}$ ({\it thin solid line}), 
the total mass of gas within 10$R_{\rm e}$ in the galaxy $M_{\rm g}$ 
({\it thick short-long dashed line}), and 
the time-integrated mass of blown-out gas $\Delta {M}_{\rm w}$ 
({\it thin short-long dashed line}). 
Only $\Delta {M}_{\rm BH}$ depends strongly on the feedback efficiency. 
Because SFR is generally 
higher than $\dot{M}_{\rm BH}$ as shown in Figure \ref{fig:MB_time1}, the increase in 
stellar mass outpaces that of the SMBH mass during the early evolution.
}
\label{fig:MB_time2}
\end{figure}

The change of the total mass in each run is summarized in Figure \ref{fig:MB_time2}. 
We do not find a strong variability in the mass of the ejected gas and gas inside a galaxy 
despite the large difference in feedback efficiencies, while we found a significant 
difference among the MA models. The main 
difference is found in the total mass of stars and the central SMBH, following the difference 
in the evolution of $\dot{M}_{\rm BH}$ and $\dot{M}_{*}$ in Figure \ref{fig:MB_time1}. 
In all simulations except for Run 1, the increase in stellar mass overwhelms the growth of the 
central SMBH before approximately 8 Gyr. As we find in the MA models, the MB model also shows that 
the dynamically evolving model of AGN feedback naturally results in the 
time-dependent mass ratio between the central SMBH and its host galaxy.

The importance of physics-based feedback models is obviously found by comparing 
Figure \ref{fig:MA_time2} and \ref{fig:MB_time2}. Even for the same radiation conversion 
efficiency from the accretion mass, the MA model is more effective than the MB model 
in supplying feedback energy and suppressing the growth of both stellar and 
SMBH mass. Although the maximum instantaneous feedback effect is same for 
the same $\epsilon_{\rm w}^{\rm M}$ in both models, 
the higher feedback effect in the MA models at low accretion rates leads to quite 
important differences. 
This difference finally affects how frequently intensive AGN activity and 
star formation is restored.

The impact from the differences between MA and MB models is particularly apparent for 
very low feedback efficiencies. For Run 1 with 
$\epsilon_{\rm w}^{\rm M} = 1 \times 10^{-5}$, the SMBH accretion rate is almost constant 
at the high value of $10^{0.4} M_{\odot}/yr$ in the MB model, while the rate strongly 
fluctuates up to about 8 Gyr in the MA model. The luminosity dependence of $\epsilon_{\rm w}$ in 
the MB models allows more rapid growth of the central SMBH than in the MA models. And then the 
very large value of $L_{\rm Edd}$ with the large $M_{\rm BH}$ consequently reduces $\epsilon_{\rm w}$ 
following Equation \ref{eq:eps} in the MB models. Therefore, the evolution quickly turns into 
a runaway state, while the central SMBH grows. 
Even though either very low or high values of $\epsilon_{\rm w}^{\rm M}$ does not show 
bursting activity (see Figure \ref{fig:MB_time1}), 
in Run 1 $M_{\rm BH}$ is much higher than that in Run 8 because the runaway process 
prevents bursts with increasing $M_{\rm BH}$ in Run 1, but simply high feedback efficiency 
in Run 8 prevents both the bursts of $\dot{M}_{\rm BH}$ and the increase in $M_{\rm BH}$.

\subsection{Problems of the MB models}

\begin{figure}[t!]
\plotone{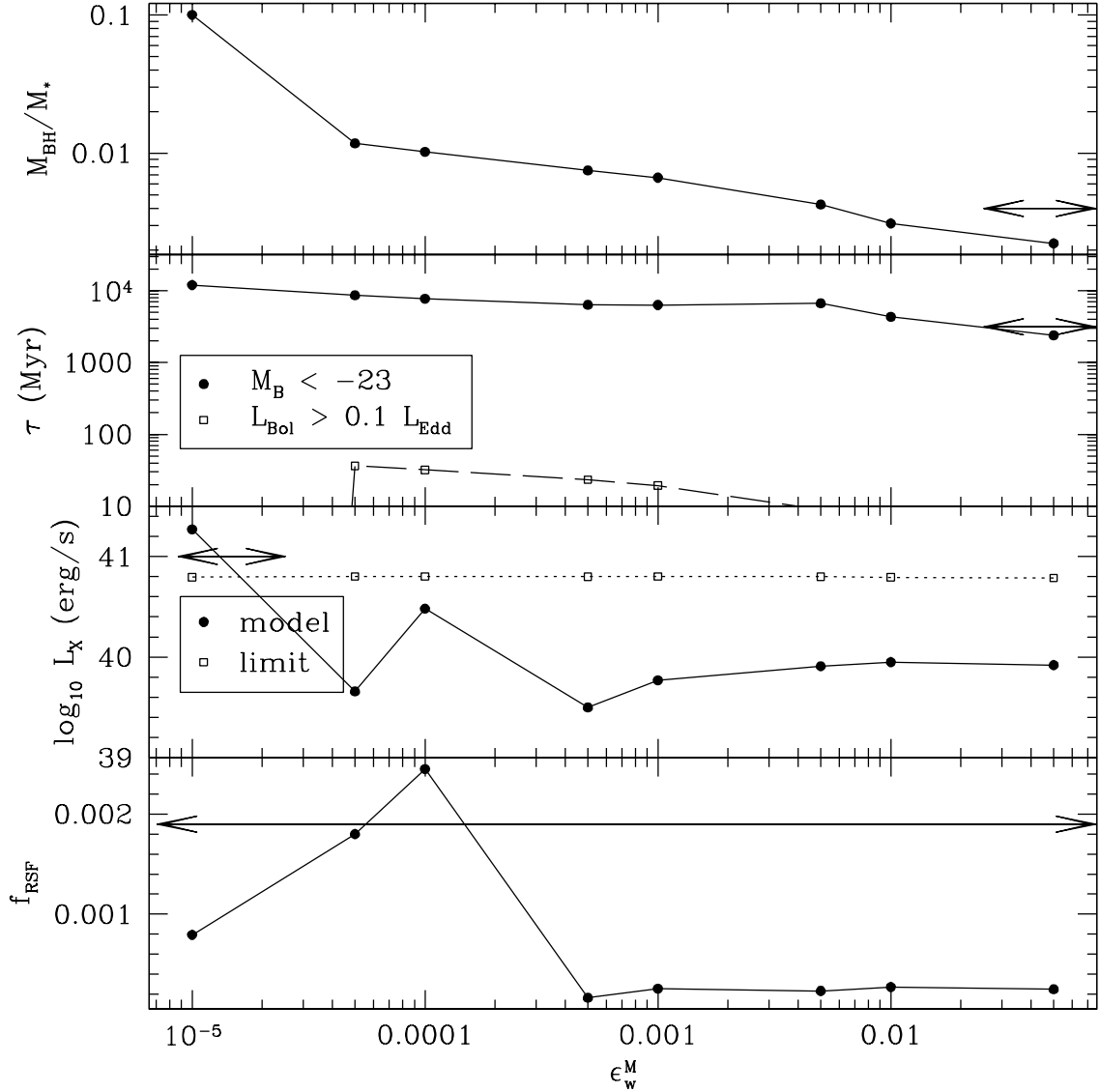}
\caption{
Results of the four tests for the MB models at the epoch of 14 Gyr. 
From top to bottom, simulation results are compared to 
four observational constraints: the SMBH mass to stellar mass ratio, 
the net quasar lifetime, the X-ray luminosity of 
diffuse hot gas, and the fraction of stellar mass formed within the last 1 Gyr. 
The low $\epsilon_{\rm w}^{\rm M}$ in the MB models enhances the growth of the 
central SMBH, resulting in the high values of $M_{\rm BH}/M_{*}$ and net quasar lifetime. 
The arrow bars represent the range of $\epsilon_{\rm w}^{\rm M}$ that is consistent 
with observations. This comparison shows 
that none of the tested feedback efficiencies can pass all tests simultaneously.
}
\label{fig:MB_comp}
\end{figure}

As we find in the tests with the MA models, the family of the MB models also appears to 
fail to pass our tests. 
Low feedback efficiencies produce too massive central SMBHs and too long net lifetime 
of quasar activity as presented in Figure \ref{fig:MB_comp}, 
while the ratio $M_{\rm BH}/M_{*}$ is much higher 
than the locally found value of about $10^{-3}$. 
Moreover, the net lifetime of quasar phase at $M_{\rm B} < -23$ mag 
is not matched to the limit of 1 Gyr \citep{martini04}. As found in the MA model, the recently 
formed stellar mass  in the MB models is also 
much lower than the observational maximum limit. 
However, the thermal X-ray luminosity for any value of $\epsilon_{\rm w}^{\rm M}$ except for 
Run 1, does not conform to the observational limits.

\begin{figure}[t!]
\plotone{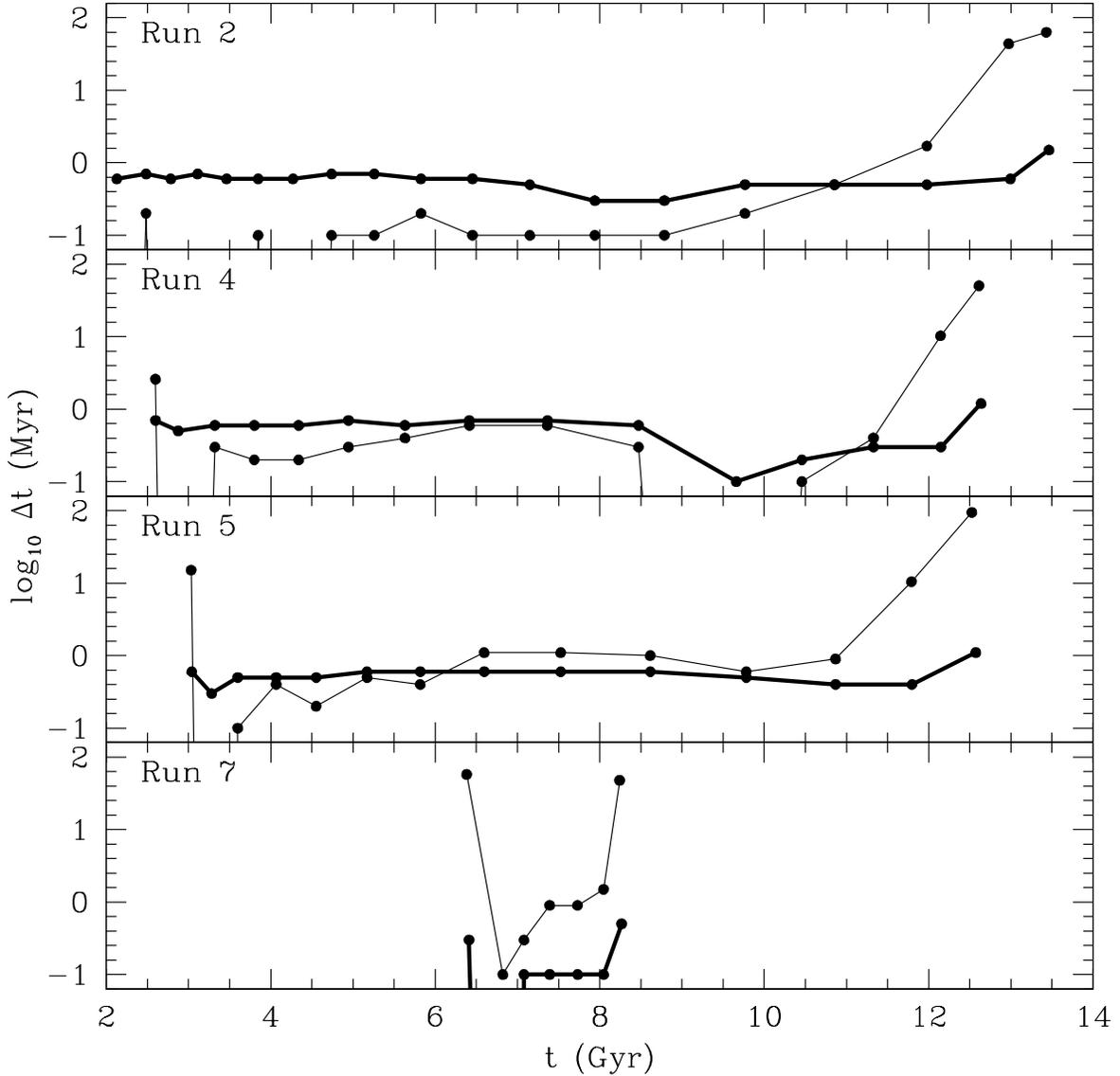}
\caption{
Episodic lifetime of intensive star formation and quasar activity in the MB models. 
For the same limits of SFR and the bolometric luminosity from the central SMBH 
used in Figure \ref{fig:MA_dT}, the MB model also shows the same pattern as the 
MA model shows. Late bursts of star formation ({\it thin line}) are 
maintained longer than early bursts. 
The duration of $L_{\rm BH} > L_{\rm Edd}$ ({\it thick line}) is between 0.1 Myr to 1 Myr generally without a 
significant dependence on $\epsilon_{\rm w}^{\rm M}$.
}
\label{fig:MB_dT}
\end{figure}

The episodic lifetime of quasar activity and significant star formation does not have 
a systematic difference in the MB models compared to the MA models. 
The duration of the last burst is considerably longer than those at earlier 
times as shown in Figure \ref{fig:MB_dT}. The first 
star burst is also longer than other peaks of SFR in the models with 
low feedback efficiencies such as Run 7. The duration of the bright 
quasar phase weakly changes considering the limit of $L_{\rm Bol} > L_{\rm Edd}$. But we note that 
the duration of episodic activity is subject to how we define the limit of activity.

The family of the MB models also fails to conform to four major observational 
constraints considered together. 
The fraction of recently formed stellar mass does not play a key role as an important 
diagnosis because of its poor sensitivity to variation of the feedback efficiency. 
The net lifetime of quasar phase is strongly coupled to the change of the SMBH mass accretion, as 
we find in the MA models. 
The X-ray luminosity of hot ISM turns out to be a useful diagnosis to test 
the feedback physics again in the MB models. The X-ray luminosity favors 
$\epsilon_{\rm w}^{\rm M} < 5 \times 10^{-5}$. But this range is much lower than 
$\epsilon_{\rm w}^{\rm M} > 10^{-2}$ which produces the right ratio between the SMBH and its 
host galaxy mass.

\section{Other possible diagnostics}

In addition to four tests adopted in this paper, we find additional possible 
diagnostics which have not been well investigated observationally, but which 
can be valuable to constrain various AGN feedback models. In particular, the feedback 
efficiencies need to be refined more precisely even in the purely mechanical feedback models 
because of the possibility that the feedback efficiency may 
be dependent of the SMBH mass and other 
properties \citep[e.g.][]{merloni08}. Therefore, finding an effective tool of 
diagnosis is quite important.

\begin{figure}[t!]
\plotone{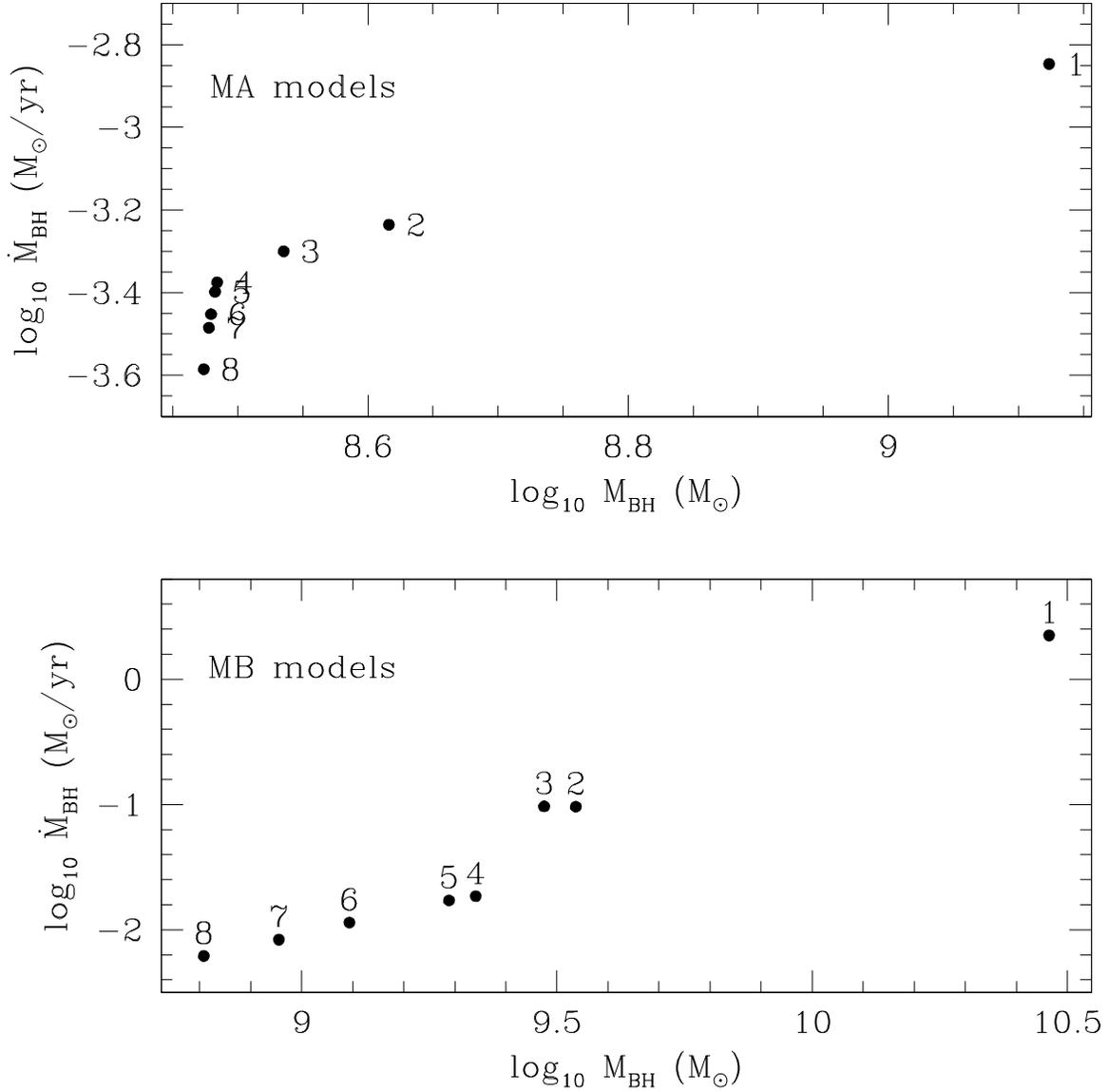}
\caption{
Distribution of the SMBH accretion rate versus its mass at 14 Gyr 
for all models in Table \ref{tab:epsilon}. The 
numbers near the points are the names of the simulation runs. 
The difference between MA and MB models results is apparent in 
the different ranges of 
$M_{\rm BH}$ for the same initial SMBH mass. In both models, high accretion rates 
correspond to low feedback efficiencies. But we note that the central SMBHs in 
all models are not in an active phase at 14 Gyr as shown in Figures \ref{fig:MA_time1} and 
\ref{fig:MB_time1}.
}
\label{fig:MA_MB_accretion}
\end{figure}

The current ratio of the SMBH accretion rate to the SMBH mass can be a constraint on 
the feedback efficiencies for different feedback models. As shown in 
Figures \ref{fig:MA_time1}, \ref{fig:MB_time1}, and 
\ref{fig:MA_MB_accretion}, the cumulative effects of 
the self-regulation process result in the difference in the SMBH mass and 
the current accretion rate \citep[see][for a discussion]{netzer07}. 
This ratio $\dot{M}_{\rm BH} / M_{\rm BH}$ can be 
converted to the observable quantity $L_{\rm BH} / L_{\rm Edd}$ which are lower than 
$10^{-3}$ in local low-luminosity AGNs and 0.1 to 1 in classical luminous local 
AGNs \citep{ho08}. As we find in Figures \ref{fig:MA_dT} and 
\ref{fig:MB_dT}, this ratio changes quickly within 1 Myr even though the 
duration of $L_{\rm BH} / L_{\rm Edd} > 1$ is quite insensitive to the feedback 
models and their efficiencies. The main concern in observation is 
measuring the statistical distribution of this ratio for broad ranges of 
AGN activity levels as well as a quite phase \citep{kim08}.

\begin{figure}[t!]
\plotone{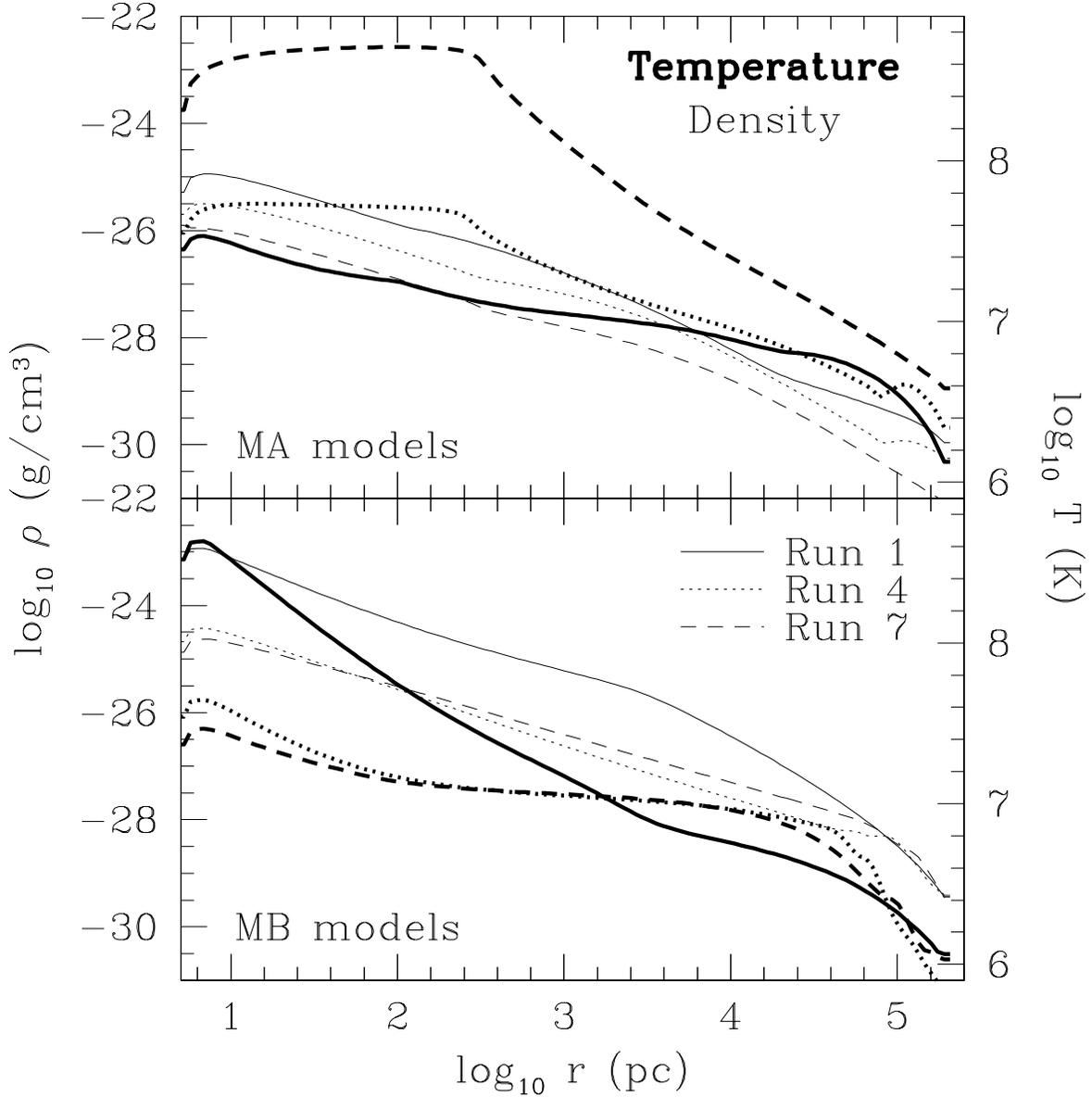}
\caption{Radial temperature and density distribution of hot gas in Run 1, 4, and 7 
at 14 Gyr when the SMBH accretion rate is much lower than the Eddington accretion rate. 
In both MA 
and MB models, the high feedback efficiencies result in the low central density ({\it thin line}). 
However, the temperature profile ({\it thick line}) 
is dependent of which feedback prescription is used. In particular, 
the temperature core in the MA models 
is produced by the constant feedback efficiency, while in MB models the low accretion 
luminosity at 14 Gyr produces a very weak feedback which makes the temperature profile be similar to 
a standard low-luminosity hot accretion profile.
}
\label{fig:MA_MB_radial}
\end{figure}

The central density and temperature of hot ISM is also an interesting quantity of the 
feedback model \citep[e.g.][]{pellegrini09}. 
The simulations show that the total amount of gas and its 
temperature change responding to the energy input by different AGN feedback models. 
In addition to this global change, the change of hot ISM in the central region 
around the SMBH shown in Figure 
\ref{fig:MA_MB_radial} 
can be compared to observations for the same range of the central SMBH and host galaxy mass. 
For example, the observed central density of electrons in a few local quiescent 
ellipticals is about 0.02 $cm^{-3}$ which corresponds to the mass density 
$\rho ~\sim~ 10^{-26} g / cm^{3}$ \citep{humphrey06,soria06a,soria06b}, which 
is not very different from the values presented in Figure \ref{fig:MA_MB_radial}. But 
as the observational values also depend on the specific galaxy models adopted, 
proper comparisons can be obtained by focusing on a well observed galaxy for which 
the dynamical and structural properties are well constrained by independent studies, or by 
studying statistically a large set of observed profiles of density and temperature for 
the X-ray emitting gas. Because we do not include recent accretion of external gas or 
major/minor galaxy mergers, a direct comparison of our simulations results to observations 
need to be limited to local ellipticals not showing dynamical and structural properties of 
recent accretion or mergers. Moreover, 
the best comparison between our simulations and observations is possible 
only with the simultaneous measurements of the SMBH mass and its accretion rate too.

\section{Discussion and conclusion}

We have shown that the implementations of two simple classes of purely mechanical feedback 
models are not likely to satisfy the four major observational constraints when considered 
together. Importantly, there does not seem to be any 
possible range of the mechanical feedback efficiencies, 
including the commonly used $\epsilon_{\rm w}^{\rm M} = 0.005$, 
which can be compatible with the observations, 
The simulation results also prove that the self-regulation process by AGNs requires 
a careful consideration of both the {\it timing problem} and of the {\it energy conversion problem}. 
As we will discuss in a subsequent paper, the models including 
both mechanical and radiative feedback may be a right approach to simulate the 
self-regulation process as we observe in local luminous quasars.

The simulations presented in this paper do not consider departures from spherical symmetry, nor 
chemical evolution of stellar and gas metallicity 
in addition to other shortcomings such as the effects of external gas pressure and accretion 
of stars to SMBH \citep[e.g.][]{jogee06,pellegrini09}. 
Because of these missing parts with the present approach, it is not possible to exploit other 
well-derived observational constraints on the feedback models. For instance, 
the commonly observed extended emission line regions around quasars are mainly understood as 
the consequence of geometrically complex outflows driven by either radiative or mechanical wind during 
the intensive AGN phase 
\citep{batcheldor07,letawe08,fu09}. Testing simulations based on the properties of the outflows 
will require one to resolve three-dimensional spatial structure, as 
various hydrodynamic or radiative instabilities are not caught in one-dimensional simulations. 
We note, however, that the adopted prescriptions for the mechanical AGN feedback effects are based 
on sub-grid physics corresponding 
to mass, momentum, and energy transportation in small-scale turbulent motions, even though the 
simulations assume a spherical symmetry. 
Implementing chemical evolution models in our simulations would be a necessary step to use 
the observed metallicity gradients and average metallicities 
of elliptical galaxies as another diagnostic \citep[e.g.][]{carollo93,gibson96}. 
Even though the effects from AGN feedback on galactic metallicity gradients have not been 
carefully considered in the current observational investigations \citep{sanchez07}, this assumption 
needs to be verified in the AGN feedback models with chemical evolution by comparing 
simulation results with observed gradients 
\citep{matteucci93,friaca98,angeletti03,ballero08,kisaka08,spolaor09}.

In our simulations, we do not consider the problem of the common evolution of the SMBH mass and 
of the stellar mass during the initial phases of galaxy formation. The simulations 
begin with a galaxy already formed and a central SMBH. 
In particular, here we assume that 
the initial SMBH may have formed with the right mass ratio for the initial bulge 
\citep[e.g.][]{elmegreen08}. But the 
establishment of the SMBH-to-stellar mass ratio, 
as well as the Faber-Jackson and the Fundamental Plane in the early evolution of ellipticals, 
is still poorly understood both observationally and theoretically \citep[see][for a review]
{ciotti09review}. It will be interesting to test different initial central black hole to stellar mass ratios 
in future simulations. Because stellar mass determines the total amount of 
recycled gas, and AGN feedback models depend on SMBH mass, probably, different initial mass 
ratios and masses might result in complicated evolution which need to be carefully tested in 
simulations.

Minor mergers and cosmological gas accretion might not affect our main conclusions. 
As we explained in the introduction, major galaxy mergers might not be a significant effect 
on late galaxy evolution. But late minor mergers and accretion can play an important role in 
fueling gas onto SMBHs. Even in these cases, our main conclusions might not be strongly changed 
except for the test with the net lifetime of the luminous quasar phase. The net lifetime 
of the quasar phase depends on how many times strong gas fueling occurs, while other 
tests such as the simultaneous growth of SMBH and stellar mass and the episodic lifetime 
of the quasar phase is dominated by a feedback physics. Cosmological simulations including 
detailed AGN feedback effects such as our models will be a direct test of effects from 
minor mergers and gas accretion.

Despite their intrinsic limitations, 
our simulations 
bridge the gap between simple prescriptions of AGN feedback in 
cosmological studies or simulations of galaxy mergers, and examination of the small-scale 
physics around the central SMBHs. The problem of the self-regulation process itself 
demands elaborate prescriptions of dynamical evolution and energy conversion around the 
central SMBH and its surrounding ISM. In this paper, we found additional evidence that 
purely mechanical feedback models, 
including some improved versions based on physical arguments, 
fail to pass basic observational constraints. 
Our next paper will 
present how combining both radiative and mechanical AGN feedback effects can produce 
the evolutionary models of elliptical galaxies that are more consistent with the 
properties of the central SMBHs and their host galaxies in local universe.

\acknowledgments

We are grateful to Michael Strauss, James Gunn, Gillian Knapp, Renyue Cen, and Christy Tremonti 
for useful discussions and careful 
reading. We thank the anonymous referee for considered comments which improved this 
manuscript. 
M.-S. is supported by the Charlotte Elizabeth Procter Fellowship of Princeton
University. Computations were performed on the computational facilities of 
PICSciE (Princeton Institute for Computational Science and Engineering).


\appendix

\section{Input physics}

In this Appendix we summarize the implementation of the physics
involved in the simulations. For a more extensive discussion the
reader is referred to \citet{ciotti07} and to Paper I.

\subsection{The hydrodynamical equations}

The evolution of the galactic gas flow is obtained
integrating the time--dependent Eulerian equations of hydrodynamics:
\beq
{\pd \rho t}+\nabla \cdot (\rho v)=\alpha\rhos +\drhoII -\drhosp,
\label{eqh1}
\eeq
\beq
{\pd m t}+\nabla\cdot (mv)=-(\gamma-1)\nabla E -\nabla\prad +g\rho -\dot m^+_*,
\label{eqh2}
\eeq
\begin{eqnarray}
{\pd Et}+\nabla \cdot (Ev)&=&-(\gamma-1)\,E\nabla \cdot v + H - C +
\label{eqh3}\\ \nonumber
&&{(\alpha\rhos +\drhoII)(v^2+3\sigast^2)\over 2} +\dEI+\dEII -\dot E^+_*.
\end{eqnarray}
$\rho$, $m$, and $E$ are the gas mass, momentum and internal energy
per unit volume, respectively, and $v$ is the gas velocity.  The ratio
of the specific heats is $\gamma =5/3$, and $g(r)$ is the
gravitational field of the galaxy (stars and dark matter), plus the
contribution of the central SMBH.  The gravitational field is updated
at each time step by considering the SMBH mass growth; for simplicity,
we do not take into account neither the ISM contribution, nor the mass
redistribution due to the stellar mass losses and star formation.  The
total radiative pressure gradient is $\nabla\prad=(\nabla\prad)_{\rm
es}+(\nabla\prad)_{\rm dust}+ (\nabla\prad)_{\rm photo}$,
while $H-C$ is the radiative heating and cooling term.

The energy source term is obtained under the assumption that the
streaming velocity of the source distribution is zero, neglecting the
small contributions of the internal energy of the injected gas, and of
the kinetic energy of stellar wind when compared to the local stellar
velocity dispersion contribution (for the derivation and detailed
discussion of the hydrodynamical equations with moving isotropic or
anisotropic source terms, see \citet{dercole00}).  The
source terms $\alpha\rhos$ and $\dEI$ of the initial, passively
evolving stellar population, and the source terms due to Type II
Supernovae, $\drhoII$ and $\dEII$, are described in the following.

The first active grid point 
$R_1$ is placed within the Compton radius
\beq
\rx={2G\mbh \mu\mpr\over 3\kb\tx}\simeq 3.6\mu {\mbh\over 10^8\Msun}
    {10^7 {\rm K}\over\tx}\quad {\rm pc},
\label{rx}
\eeq
so that at $R_1$ we can impose the physical condition of a vanishing
{\it thermodynamical} pressure gradient, leading to gas free-fall on
the circumnuclear disk when the radiation pressure is negligible; in
this paper we adopt $\tx= 2.5\times 10^7$ K.  The appropriate values
for radiation pressure at $R_1$ are obtained from the disk treatment.

The simulations are realized with a spatially second-order Eulerian
scheme which adopts two staggered grids, each of them consisting of
120 logarithmically spaced grid points.  The equations are integrated
with a time-splitting scheme, while the heating and cooling terms in
the energy equation are integrated by using a predictor-corrector
scheme, so that the integration is second order in time. At each
simulation time, the time-step is determined as a fraction of the
minimum among the Courant condition over the grid, and of the others
characteristic times associated with the described physical processes:
during the accretion phases (and subsequents bursts of radiation), it
is not infrequent to have time-steps of the order of 1 yr or
less. However, it is important to note the accretion events are
characterized by the intrinsic time-scale related to
equation~(\ref{rx}) by
\beq
t_{\rm X}\equiv {\rx\over c_{\rm X}}\simeq 1.22\,10^4 \mu^{3/2} 
                {\mbh\over 10^8\Msun}
                \left({10^7 {\rm K}\over\tx}\right)^{3/2}\quad {\rm yr},
\label{timex}
\eeq
where $c_{\rm X}$ is the isothermal sound velocity associated with the 
Compton temperature.

\subsection{Stellar passive evolution: SNIa rate and stellar mass losses}

The stellar mass loss rate and the SNIa rate associated with the
initial stellar distribution are the main ingredients driving
evolution of the models. In the code the stellar mass losses -- the
source of {\it fuel} for the activity of the SMBH -- follow the detailed
prescriptions of the stellar evolution theory.
Over the whole galaxy
\beq
\dot\mast={\rm IMF}(\MTO)|\dot\MTO|\Delta M,
\eeq
where the initial mass function IMF is a Salpeter law (normalized as
described in CDPR), and the turn-off mass (in $\Msun$) of stars at time 
$t$ (in Gyrs) is
\beq
\log \MTO=0.0558(\log t)^2-1.338\log t +7.764.
\eeq
Finally
\beq
\Delta M=\cases{\MTO-M_{\rm fin}(\MTO)=0.945\MTO-0.503,\quad (\MTO< 9\Msun),\cr
                \Delta M=\MTO-1.4\Msun,\quad (\MTO\geq 9\Msun).}
\eeq

The time evolution of the SNIa rate is parametrized as
\beq
\Rsn (t)=0.32\times 10^{-12}h^2\vtsn {\lb\over L_{\rm B\odot}}
          \left ({t\over 13.7\,{\rm Gyr}}\right)^{-s}\quad {\rm yr}^{-1},
\eeq
where 
$h\equiv H_{\circ}/100$ km s$^{-1}$ Mpc$^{-1}$, and 
the coefficient $\vtsn$ allow for different choices in the
present-day SNIa.  Assuming for each supernova event an energy release of
$\Esn=10^{51}$ erg, a fraction $\etaII$ of which is thermalized in the
surrounding ISM, the energy input per unit time over all the galaxy
body is given by
\beq
\lsn (t)=1.015\times 10^{31}h^2\vtsn\etaII {\lb\over L_{\rm B\odot}}
          \left ({t\over 13.7\,{\rm Gyr}}\right)^{-s}
          \quad\quad {\rm erg}\,{\rm s}^{-1};  
\eeq
in this paper we restrict to the case $\vtsn =1$ and $h=0.75$.  Here
we restrict to the currently favoured $s=1.1$ value.

Besides energy, supernovae provide also mass. We assume that each SNIa
ejects $1.4\Msun$ of material in the ISM, so that the total rate of
mass return from the aging initial stellar population at each place in
the galaxy is
\beq
{d\rhos\over dt}=(\als +\asn)\rhos ,
\label{alphas}
\eeq
where $\asn (t)=1.4\Msun\,\Rsn (t)/\mast$ and $\als (t)=\dot\mast (t)
/\mast$ are the specific mass return rates.  With these definitions,
the SNIa kinetic energy injection per unit volume in the ISM can be
written as
\beq
\dEI=\etaII\Esn{\Rsn\over\mast}\rhos=\etaII\Esn{\asn(t)\rhos\over 1.4\Msun}.
\label{dEI}
\eeq

\subsection{Star formation, SNII heating and starburst properties}

Star
formation cannot be avoided when cool gas accumulates in the
central regions of elliptical galaxies. In particular, we compute the
star formation rate at each radius $r$ from the equation
\beq
\drhosp={\etaform\rho\over\tform},\quad
\tform=\max (\tcool,\tdyn),
\label{drhosp}
\eeq
where $\rho$ is the local gas density, $\etaform=0.03 - 0.4$,
and the associated characteristic times are
\beq
\tcool\equiv {E\over C},\quad
\tdyn=\min(\tjeans,\trot),\quad
\tjeans\equiv\sqrt{3\over 32\pi G\rho},\quad
\trot\equiv{2\pi r\over v_c(r)}.
\eeq
$E$ and $C$ are the gas internal energy and the
effective cooling per unit volume, while
$v_c(r)$ is the galaxy rotational velocity at radius $r$. In the code
the stars are maintained in the place where they form, and in each
shell the associated sinks of momentum and internal energy per unit
volume are given by the negative of
\beq
\dot m_*^+={\etaform m\over\tform}, \quad 
\dot E_*^+={\etaform E\over\tform},
\eeq
where $m$ is the specific momentum of the ISM.

For a total mass $\Delta\mast$ of newly formed stars in a given
time-step and at a given place, we assume a Salpeter IMF
\beq
{dN\over dM}=
(x-1)\left({\Min\over\Msun}\right)^{x-1}{\Delta\mast\over\Msun}\times
\left({M\over\Msun}\right)^{-1-x},\quad 
(x >1, M\geq\Min =0.1\Msun),
\label{dNdM}
\eeq
so that the associated total number of Type II Supernovae is
\beq
\NII=\int_{\MII=8\Msun}^{\infty}{dN\over dM}dM=\left(1-{1\over x}\right)
\left({\Min\over\MII}\right)^x{\Msun\over\Min}{\Delta\mast\over\Msun}
\simeq 7\times 10^{-3}{\Delta\mast\over\Msun},
\eeq
where the numerical value holds for $x=1.35$.  As for SNIa, we assume
that each SNII event releases $\Esn=10^{51}$ erg of kinetic energy, 
and the resulting mean efficiency is
\beq
\epsII\equiv {\NII\Esn\etaII\over\Delta\mast c^2} = 
\left(1-{1\over x}\right)
\left({\Min\over\MII}\right)^x{\Msun\over\Min}{\Esn\etaII\over\Msun c^2}
\simeq 3.9\times 10^{-6}\etaII;
\label{epsII}
\eeq
in this paper we assume $\etaII=0.85$.  The characteristic time for
SNII explosion is fixed to $\tauII=2\times 10^7$ yr, and from
equations~(\ref{drhosp}) and (\ref{epsII}) their luminosity (per unit
volume) at each radius from the galaxy center is
\beq
\dEII (t)\equiv {\epsII c^2\over\tauII}\int_0^t\drhosp (t')
e^{-(t-t')/\tauII}dt'.
\label{dEII}
\eeq

We assume that each explosion leaves a neutron stars of $1.4\Msun$. 
As a consequence, the total mass ejected
by the SNII explosions per unit mass is
\beq
{\MII^{ej}\over\Delta\mast}=\left({\Min\over\MII}\right)^{x-1} 
                       -1.4{\NII\Msun\over\Delta\mast}\simeq 0.2,
\eeq
and the mass return rate per unit volume of the young evolving stellar
population is given by
\beq
\drhoII(t)\simeq {0.2\over\tauII}\int_0^t\drhosp (t')
e^{-(t-t')/\tauII}dt'.
\label{drhoII}
\eeq

Finally, in the code we also compute the fiducial optical and UV
luminosity per unit volume of the new stars as
\beq
\dEopt (t)\equiv {\epsopt c^2\over\tauopt}\int_0^t\drhosp (t')
e^{-(t-t')/\tauopt}dt',
\label{dEopt} 
\eeq
and 
\beq
\dEUV (t)\equiv {\epsUV c^2\over\tauUV}\int_0^t\drhosp (t')
e^{-(t-t')/\tauUV}dt',
\label{dEUV} 
\eeq
respectively, where $\epsopt =1.24\times 10^{-3}$, $\epsUV =8.65\times
10^{-5}$, $\tauopt =1.54\times 10^8$ yr, and $\tauUV =2.57\times 10^6$
yr are the efficiency and characteristic time of optical and UV
emission, respectively. 

\subsection{Radiative heating and cooling}

Compton heating and cooling, \brem losses, line
and recombination continuum heating and cooling, are taken into
account.

A good approximation to the net gas energy change rate $\dot E$, valid
for $T\gsim 10^4$ K (all quantities are expressed in cgs system) is
given by
\beq
\dot E = n^2 (S_1 + S_2 + S_3)\equiv H-C,
\label{dotE}
\eeq
where $n$ is the Hydrogen density (in number), and positive and
negative terms are grouped together in the heating ($H$) and cooling
($C$) functions. The \brem losses are given by
\beq
S_1 = -3.8\times 10^{-27}\sqrt{T},
\eeq
the Compton heating and cooling is given by
\beq
S_2 = 4.1\times 10^{-35} (\tx -T)\,\xi ,
\label{eqS2}
\eeq
where $\tx$ is the Compton temperature, and
finally the sum of photoionization heating, line and recombination
continuum cooling is
\beq
S_3 = 10^{-23}{a + b\, (\xi/\xi_0)^c\over 1 + (\xi/\xi_0)^c},
\label{eqS3}
\eeq
where
\beq
a=-{18\over  e^{25 (\log T -4.35)^2}} 
    -{80\over  e^{5.5(\log T -5.2)^2}}
    -{17\over  e^{3.6(\log T -6.5)^2}},
\eeq
\beq
b=1.7\times 10^4\;T^{-0.7}, 
\eeq
\beq
c=1.1-{1.1\over  e^{T/1.8\,10^5}}+{4\times 10^{15}\over T^4}, 
\eeq 
and 
\begin{eqnarray}
\xi_0 &=& {1\over 1.5\; T^{-0.5}+1.5\times 10^{12}\; T^{-2.5}}+\nonumber\\
      &&  {4\times 10^{10}\over T^2}
          \left[1 + {80\over e^{(T-10^4)/1.5\,10^3}}\right].
\end{eqnarray}
Equations~(\ref{eqS2})-(\ref{eqS3}) depend on the ionization parameter
\beq
\xi\equiv {\lbhefphot (r)\over n(r) r^2},
\eeq
where $\lbhefphot (r)$ is the effective accretion luminosity at $r$,
which is evaluated by numerically solving in each shell the balance
equation
\beq
{d\lbhefphot (r)\over dr}=-4\pi r^2 H,
\eeq
with central boundary condition $\lbhefphot (r=0)=\lbh(t)$ given by
equation~(\ref{eq:lbhbol}).  The photoionization+Compton opacity
associated with radiation absorption is then obtained
\beq
\kpho=-{1\over\rho\lbhefphot (r)}{d\lbhefphot (r)\over dr}=
       {4\pi r^2 H(r)\over\rho(r)\lbhefphot (r)}.
\eeq
Finally, the bolometric ISM luminosity is obtained from equation~(\ref{dotE})
as 
\beq
\lrad(r)=4\pi\int_0^r Cr^2 dr.
\label{lrad}
\eeq

\subsection{Radiation pressure}

Radiation pressure due to {\it electron
scattering} (where neither the photon numbers, nor their energy
change) is computed as
\beq
(\nabla\prad)_{\rm es}=-{\kes\rho\over c}
                          {\lbh+\luv(r)+\lopt(r)+\lrad (r)\over 4\pi r^2},
\label{prad_es}
\eeq
where $\kes=0.35$ in c.g.s. units, and from
equations~(\ref{dEopt})-(\ref{dEUV})
\beq
\luv(r)=4\pi\int_0^r \dEUV r^2 dr,\quad
\lopt(r)=4\pi\int_0^r \dEopt r^2 dr.
\eeq
Note that all the luminosities used in equation~(\ref{prad_es}) are
unabsorbed.

The radiation pressure contribution due to {\it dust opacity} is given
by
\beq
(\nabla\prad)_{\rm dust}=-{\kUV\rho\over c}
                          {\lbhefUV(r)  +\luveff(r)\over 4\pi r^2}
                         -{\kopt\rho\over c}
                          {\lbhefopt (r)+\lopteff(r)\over 4\pi r^2}
                         -{\kIR\rho\over c}{\lir (r)\over 4\pi r^2},
\label{prad_dust}
\eeq
where 
\beq
\lir(r)\equiv L_{\rm BH,UV}^{\rm abs}(r)+L_{\rm BH,opt}^{\rm abs}(r)+
           \luv^{\rm abs}(r) +\lopt^{\rm abs}(r), 
\label{eq_lir}
\eeq
is the infrared luminosity due to recycling of photons absorbed from
the ISM, and we adopt as estimates for (cgs) opacity in three bands
\beq
\kopt={300\over 1 + T/10^4},\quad\kUV= 4\kopt,\quad \kIR={\kopt\over 150}.
\eeq
At variance with electron scattering the {\it
effective} luminosities appearing in
equations~(\ref{prad_dust})-(\ref{eq_lir}) take into account
absorption, and are obtained by numerically solving the two lowest
spherically symmetric moment equations of radiative transfer in the
Eddington approximation (e.g., Chandrasekhar 1960):
\beq
{d\luveff\over dr}=4\pi r^2(\dEUV-\kUV\rho\Juveff),\quad
{d\lopteff\over dr}=4\pi r^2(\dEopt-\kopt\rho\Jopteff).
\label{prad_eff1}
\eeq
\beq
{d\Juveff\over dr}=-{3\kUV\rho\luveff\over 4\pi r^2},\quad
{d\Jopteff\over dr}=-{3\kopt\rho\lopteff\over 4\pi r^2},
\label{prad_eff2}
\eeq
The central boundary conditions for stellar luminosities are 
$\luveff(0)=\lduv$, $\lopteff(0)=\ldopt$, $\Juveff (0)=\lduv/16\pi^2R_1^2$ 
and $\Jopteff(0)=\ldopt/16\pi^2R_1^2$.
The effective accretion luminosities $\lbhefUV$ and $\lbhefopt$ are
computed with two equations similar to (\ref{prad_eff1}), where the
distributed source term is missing, $J=\lbhef/4\pi r^2$, and in the UV
and optical bands $\lbhefUV (0)=0.2\lbh(t)$ and $\lbhefopt
(0)=0.1\lbh(t)$, respectively.

The last contribution to radiation pressure comes from {\it
photoionization opacity},
\beq
(\nabla\prad)_{\rm photo}=-{\rho\kpho\over c}
                           {\lbhefphot (r)\over 4\pi r^2}.
\label{prad_photo}
\eeq

\subsection{The circumnuclear disk and the SMBH accretion luminosity}

The circumnuclear disk, which is the repository of the gas inflowing
at a rate $\dmineff$ from the first active mesh point $R_1$ of the
hydrodynamical grid, and which feeds the central SMBH at a rate
$\dot\mbh$, contains at any time the mass gas $\Mdg$ and a total
stellar mass $\Mds =\Mdsl +\Mdsh$, which is divided among low and high
mass stars (with the division mass at $8\Msun$).  The disk also
contains a mass $\Mrem$ of remnants from the earlier generations of
evolved stars.

In the adopted scheme the accretion rate on the central SMBH is given
by
\beq
\mdot ={\dMfid\over 1+\etaD},
\label{eq:mdotbh}
\eeq
where
\beq
\dMfid\equiv {\Mdg\over\tlagd},\quad 
\etaD\equiv {\dMfid\over 2\dot\Medd},\quad
\dot\Medd\equiv{\ledd\over\epsz c^2}
\label{eq:mfid}
\eeq 
are the fiducial depletion rate of gas from the circumnuclear disk,
its normalized value, and the Eddington mass accretion rate,
respectively. The reference radiative efficiency $\epsz$ is defined in
eq.~(\ref{eq:ADAF}).  Equations (\ref{eq:mdotbh})-(\ref{eq:mfid}) are
designed to guarantee that when $\etaD\ll 1$ the gas is accreted onto
the central SMBH at the rate $\dMfid$, while $\dot\mbh=2\dot\Medd$ for
$\etaD\gg 1$ (i.e., we allow for possible moderate super-Eddington
accretion; note however that outside the first grid point $R_1$ the
flow accretion rate is limited in a self-consistent way by feedback
effects). From eq.~(\ref{eq:mdotbh}) we calculate the instantaneous
bolometric accretion luminosity as
\beq
\lbh =\epsA\,\mdot\,c^2,
\label{eq:lbhbol}
\eeq
where
\beq
\epsA =\epsz{A\dot m\over 1+A\dot m},\quad 
\dot m\equiv {\mdot\over \dot\Medd},
\label{eq:ADAF}
\eeq
and $A$ is a free parameter so that $\epsA\sim \epsz A\dot m$ for
$\dot m \ll A^{-1}$.  In our simulations we fix $A=100$, and we introduce the
normalized accretion luminosity
\beq
l\equiv{\lbh\over\ledd}={A\dot m^2\over 1 + A\dot m},
\label{eq:lbhnor}
\eeq
where the last expression derives from the ADAF phenomenological 
description.

There are a few {\it lag times} in our problem which are expressed as follows.
The first is the {\it instantaneous disk lag time}, appearing
in eq.~(\ref{eq:mfid}): 
\beq
\tlagd\equiv{2\pi\over\alpha}\sqrt{\Rd^3\over G\mbh}, 
\label{eq:tlagd}
\eeq 
where $\alpha\simeq 10^{-2}-10^{-1}$ is the disk viscosity
coefficient, and $\Rd$ and $\mbh$ are the instantaneous values of the
fiducial radius of the circumnuclear disk and the mass of the central
SMBH. We use the scaling predicted by thin-disk theory
\beq
\Rd(t) =f_d R_1\times\left ({\mbh \over M_{\rm BH0}}\right)^{2/3},
\label{eq:rdisk}
\eeq 
where $M_{\rm BH0}$ is the central SMBH
mass at the beginning of the simulation. We assume $f_d=0.4$, so that
$\Rd (0)\simeq 2$ pc for an initial SMBH mass of $\simeq 10^8\Msun$.

The second characteristic time is the
instantaneous {\it infall} lag time from $R_1$ to the disk:
\beq
\tlagi={R_1\over\vff},\quad 
\vff\equiv\sqrt{2G\mbh\over R_1},
\label{eq:tff}
\eeq
so that the effective rate at which gas accretes on the disk is obtained 
by solving the differential equation
\beq
{d\dmineff\over dt}={\dmin -\dmineff\over\tlagi},
\label{eq:dmineff}
\eeq
where $\dmin$ is the instantaneous rate at which gas flows through the
first active grid point\footnote{$\dmin$ is taken positive in case of
  accretion and zero in case of outflow at $R_1$.}.  It follows that
when $\dmin$ provided by hydrodynamics drops to zero the circumnuclear
disk experiences a fueling declining exponentially with time.

The disk total gas mass $\Mdg$ is not only the source of SMBH
accretion, but also of star formation in the disk: we assume that a
fraction of $\Mdg$ is converted into stars at a rate $\etas\dMfid$
(where $\etas\simeq 10\Mdg/\mbh$), and that another fraction of $\Mdg$
is lost as a {\it disk wind} and as a {\it jet} at instantaneous rates given by
$\etaw\dot\mbh$ and $\etaj\dot\mbh$, so that the equation for the gas
mass in the disk is
\beq
{d\Mdg\over dt}=\dmineff - (1+\etaw+\etaj)\dot\mbh -\etas\dMfid.
\label{eq:dmdg}
\eeq 
The stars formed in the disk are described
separately as a function of their mass, i.e., high-mass stars ($M>\MII
=8\Msun$) produce a total disk mass $\Mdsh$, and low-mass stars ($\Min
<M<\MII$) contribute to a disk mass $\Mdsl$ according to the equations
\beq
{d\Mdsl\over dt}=(1-\fh)\etas\dMfid -{\Mdsl\over\taul};\quad
{d\Mdsh\over dt}=\fh\etas\dMfid -{\Mdsh\over\tauh}.
\label{eq:dmdstar}
\eeq
For the characteristic evolutionary times we adopt $\taul
=\tauopt$ and $\tauh=\tauII$, while we assume $\fh =
0.5$, corresponding to a top-heavy Salpeter-like initial mass function
of
slope $x\simeq 1.16$ and minimum mass $\Min = 0.1\Msun$.  The
associated optical ($\ldopt$) and UV ($\lduv$) luminosities of the
stellar disk are calculated following the scheme described in \citet{ciotti07}.
Finally stellar remnants mass in the disk evolves as
\beq
{d\Mrem\over dt}=\freml{\Mdsl\over\taul}+\fremh{\Mdsh\over\tauh},
\label{eq:dmdrem}
\eeq
where $\freml=0.2$, $\fremh=0.09$. 

The equation for the mass loss associated with the disk wind is 
\beq
{d\Mdw\over dt}=\etaw\mdot+
               (1-\freml){\Mdsl\over\taul}+
               (1-\fremh){\Mdsh\over\tauh}:
\label{eq:mdiskw}
\eeq
the first term is a mass loss driven as a wind by the central SMBH, and the
second and third are from high mass and low mass stars in the central
disk.

We explore two different classes of models, that we call {\it Type A}
and {\it Type B}, with 
\beq
\etaw\equiv\cases{2,\quad\quad\quad\quad\quad\quad\quad\quad\,\; {\rm
    [A]}\cr \displaystyle{{3\etawM\over 4}{l\over 1+0.25 l}},
  \quad\quad\quad {\rm [B].}}
\label{eq:etaw_appendix}
\eeq

We also consider another mass 
component ejected by disk, i.e. a {\it nuclear jet} with instantaneous 
mass flow
\beq
{d\Mj\over dt}=\etaj\mdot,\quad\etaj={0.2\over (1+100l)^4},
\label{eq:dmj}
\eeq
so that the mass ejected by the jet is always negligible with respect
to the wind mass loss in Type A models, while it is slightly dominant
over the wind in Type B models at low luminosity ratios.  In the code,
all the equations presented in this Section are integrated numerically
with a first order finite difference scheme.

\subsection{The mechanical feedback treatment}

We now discuss how the kinetic energy, momentum and mass of the BLR
wind are transferred to the ISM. 
The fiducial {\it instantaneous mechanical luminosity} of the 
disk wind is given by
\beq
\ldwin =\epsw\dot\mbh c^2 +\epsII c^2 (1-\fremh){\Mdsh\over\tauh},
\label{eq:ldwin}
\eeq
where $\epsw$ is the mechanical efficiency of the wind, and the second
term describes the energetic associated with the SNII explosions of
the high-mass stars in the circumnuclear disk. In analogy with
eq.~(\ref{eq:etaw_appendix}), we assume
\beq
\epsw\equiv\cases{\epswM,\quad\quad\quad\quad\quad\quad\quad\;\;\, {\rm [A]}\cr
                  \displaystyle{{3\epswM\over 4}{l\over 1+0.25 l}},
\quad\quad\quad {\rm [B].}}
\label{eq:epsw}
\eeq
In Type A models we explore the range $3\,10^{-5}\leq\epswM\leq
5\,10^{-3}$. In Type B models, where the wind efficiency is a function of the
normalized accretion luminosity, $\epswM$ is the maximum possible
value (reached for $l=2$). In both cases the {\it
  instantaneous disk wind velocity} is given by
\beq
\vw\equiv\sqrt{2\ldwin\over\dot\Mdw}\simeq\sqrt{2\epsw\over\etaw}c,
\label{eq:vdiskw}
\eeq
where the last expression neglects the mass return contribution of
massive stars in the circumnuclear disk. In Type A models $\vw$ is in the range $2\times10^3 - 2\times10^4$
km s$^{-1}$ (as a function of the specific assumed value for
$\epsw$), in agreement with observations of BLRs. 
For the same reasons, in Type B models we require
$\vw=10^4$ km s$^{-1}$, so that $\etawM$ and $\epswM$ are linked by
the relation
\beq
\etawM=1800\epswM.
\label{eq:modB}
\eeq

In analogy with the wind component, the {\it instantaneous jet 
mechanical luminosity} is written as 
\beq
\lj =\epsj\mdot c^2,\quad \epsj={0.0125\over (1+400l)^4},
\label{eq:ldjet}
\eeq
and the jet velocity is given by
\beq
\vj\equiv\sqrt{2\lj\over\dot\Mj}=\sqrt{2\epsj\over\etaj}\,c,
\label{eq:vdiskj}
\eeq
which, for our chosen parameterization gives high but subrelativistic
jet velocity of $\vj/c\simeq 10^{-1.65}$ for $l\gsim 0.1$.
Finally, the wind and jet momentum are defined as
\beq
\mj\equiv\dot\Mj\vj;\quad \mw\equiv\dot\Mdw\vw. 
\label{eq:momwj}
\eeq
We now illustrate how we distribute the mechanical feedback over the
galaxy ISM.
First we introduce the
{\it instantaneous wind and jet lag times}
\beq
\tau_{\rm wj}\equiv{R_1\over \vwj}
\label{eq:tlagw}
\eeq 
from the center to the first active grid point $R_1$ (where the
subscript indicates the specific component - disk wind or nuclear jet
- considered), and at each time step we compute the time-lagged values
for mass, momentum, and kinetic energy at $R_1$ by solving the
differential equation
\beq
{dX_l\over dt}={X-X_l\over\tau_{\rm wj}},
\eeq
where $X_l$ is the generic lagged variable associated with the instantaneous
unlagged value $X$.  Outside $R_1$ we then distribute
mass, momentum and kinetic energy over the hydrodynamical grid 
(outside $R_1$), by integrating numerically the phenomenological differential
equation
\beq
{\partial\ln\Ywj\over \partial\ln r}=-{\Pism (r)\over\Pwj (r)}-
                            {r\over \vwj}
                            {\partial\ln\Ywj\over\partial t},
\label{eq:mechfed}
\eeq
where $\Ywj$ is the mass, momentum and energy of the disk wind/jet
component at distance $r$ from the center, $\Pwj(r)$ is the local
wind/jet pressure, and for each quantity $Y(R_1)=X_l$. In this paper
we restrict to simulations where the time derivative is neglected.
In practice, we first integrate eq.~(\ref{eq:mechfed})
for the wind/jet pressure, i.e.,
\beq
\Pwj={\Ywj\over 2\Delta\Omega_{\rm wj} r^2},
\label{eq:pwj}
\eeq
where $\Ywj$ is the effective wind/jet momentum crossing the shell of
radius $r$, so that eq.~(\ref{eq:mechfed}) is a non-linear
differential equation for $\Ywj$. Once the equation is integrated, the
radial behavior of $\Pwj$ and the r.h.s.  of eq.~(\ref{eq:mechfed})
are known over the whole grid, and the equation can be integrated for
mass and energy.

The solid angle in the denominator of eq.~(\ref{eq:pwj}) is the
opening angle of the wind and of the jet, and the factor of 2 accounts
for the biconical nature of the flow. While for the jet we
assumed in all the simulations the fiducial value
$\DOmej=2.5\;10^{-2}$, for the wind case we adopt
\beq
\DOmew=\cases{
\pi\quad\quad\quad\quad\quad\quad\quad\quad\quad\quad\;\;\;\,{\rm [A]}\cr
\pi\min(\sqrt{l^2+a^2},1),\quad\quad\quad\;{\rm [B]},}
\label{eq:domeg}
\eeq 
where case B is designed to mimic the behavior found in radiation
driven winds: higher luminosity corresponds to a larger opening
angles. The constant inside the square root is fixed to
$a=\DOmej/\pi$, so that for small values of accretion luminosity the
wind opening angle coincides with the jet opening angle. Finally, note
that the almost linear dependence of $\DOmew$ on $l$ for $l>a$ assumes
that the {\it linear} opening angle depends on $\sqrt{l}$ for this
regime. 

To implement numerically the mechanical feedback terms, we finally
compute the nuclear wind mass, momentum and kinetic energy per unit
volume deposited in each shell as
\beq
{\rm Source}_{\rm wj}={3\over 4\pi}{\Ywj (R_i)-\Ywj (R_{i+1})\over 
                       R_{i+1}^3-R_i^3}
\eeq
and we add them (only for the wind component) to the r.h.s.  of
eqs.~(\ref{eqh1})-(A3).



\begin{table}[!t]
\caption{Efficiency parameter of the mechanical feedback in the both MA and MB 
computed models.}
\begin{tabular}{cc|cc} \hline \hline
Run & $\epsilon_{\rm w}^{\rm M}$ & Run & $\epsilon_{\rm w}^{\rm M}$\\ \hline
1 & $1\times10^{-5}$ & 5 & $1\times10^{-3}$ \\
2 & $5\times10^{-5}$ & 6 & $5\times10^{-3}$ \\
3 & $1\times10^{-4}$ & 7 & $1\times10^{-2}$ \\
4 & $5\times10^{-4}$ & 8 & $5\times10^{-2}$
\end{tabular}
\label{tab:epsilon}
\end{table}



\end{document}